\documentclass[12pt,a4paper]{article}
\usepackage[a4paper,margin=26mm]{geometry}
\usepackage{fontspec}
\usepackage{polyglossia}
\setmainlanguage{urdu}
\setotherlanguage{english}
\newfontfamily\urdufont[
  Script=Arabic,
  Scale=1.25,
  UprightFont=Amiri-Regular.ttf,
  BoldFont=Amiri-Bold.ttf,
  ItalicFont=Amiri-Italic.ttf,
  BoldItalicFont=Amiri-BoldItalic.ttf
]{Amiri-Regular.ttf}
\newfontfamily\englishfont{lmroman10-regular.otf}
\usepackage{amsmath,amssymb,amsthm,mathtools}
\usepackage{graphicx}
\usepackage{hyperref}
\hypersetup{colorlinks=true,linkcolor=blue,citecolor=blue,urlcolor=blue}
\usepackage{enumitem}
\newcommand{\justify}{}

\begin{document}


\title{کوانٹم نظریۂ تدبیر اور ارتقائی استحکام: اصول، نتائج اور کھلے سوالات}
\author{اظہر اقبال \quad اور \quad ڈیرک ایبٹ\\[3pt]
\small \textenglish{Azhar Iqbal and Derek Abbott}\\
\small \textenglish{School of Electrical and Mechanical Engineering, Adelaide University, Australia}}
\date{}

\begin{center}\Large\textbf{کوانٹم نظریۂ تدبیر اور ارتقائی استحکام: اصول، نتائج اور کھلے سوالات}\\[8pt]\normalsize اظہر اقبال اور ڈیرک ایبٹ\\[4pt]\textenglish{Azhar Iqbal and Derek Abbott}\\\textenglish{Adelaide University, Australia}\end{center}
\section*{خلاصہ}
اس نظر ثانی شدہ اور مفصل جائزہ مضمون میں کلاسیکی نظریۂ تدبیر سے کوانٹم کھیلوں اور ارتقائی استحکام تک کے فکری اور ریاضیاتی سفر کو اردو میں بیان کیا گیا ہے۔ نئے ابواب میں کیوبٹ، پیمائش، انٹینگلمنٹ اور شور کی بنیادی تشریح، قیدیوں کے مخمصے اور باز--فاختہ کھیل کے قدم بہ قدم حساب، ایک کوانٹم طریقۂ کار کی تعمیر، کئی کھلاڑیوں والے کھیل، اردو اصطلاحات کی فرہنگ، اور خود جانچ کے سوالات شامل ہیں۔ نیش کے توازن، قیدیوں کے مخمصے، مخلوط تدابیر اور ارتقائی طور پر مستحکم تدبیر کی تعریفیں مثالوں کے ذریعے واضح کی گئی ہیں۔ آیزرٹ، ولکنز اور لیونسٹائن کے کوانٹم کھیل کی تشریح کے ساتھ یہ بنیادی احتیاط بھی پیش کی گئی ہے کہ کسی کوانٹم توازن کا وجود اور اس کا استحکام اجازت یافتہ تدابیر، پیمائش، اور باہم مساوی ادائیگی رکھنے والی تدابیر کی شناخت پر منحصر ہیں۔ آخر میں تحقیق کے کھلے سوالات اور اس میدان کی حدود اور امکانات پر گفتگو کی گئی ہے۔

\begin{english}
\textbf{Abstract.} This expanded Urdu-language review introduces classical and quantum game theory with particular attention to Nash equilibrium and evolutionary stability. It adds worked examples of the Prisoner's Dilemma and Hawk--Dove games, a concise introduction to qubits and measurement, quantum protocol construction, noise robustness, multiplayer games, a bilingual glossary, and guided questions for learners. It explains the role of quantum protocols, entanglement, admissible strategy spaces, and payoff equivalence, while distinguishing mathematical game-theoretic conclusions from claims of physical implementation. Detailed new sections distinguish equilibrium from ESS and from stability of payoff-equivalence classes, derive the instability of a classical mixed coordination equilibrium, discuss the Bloch-sphere quotient for a special quantum game, analyze noise and perturbations, and propose a reproducible research framework. The article ends with limitations and open research questions.
\end{english}

\justify

\section{انتساب}
       يہ مضمون پاکستان کے ان طلبہ و طالبات کے نام وقف ہے جو ذہنی تجسّس اور علمی جذبے کے ساتھ  ساتھ اردو سے شغف بھی رکھتے ہيں۔

\newpage
\section{تعارف}
گیم تھیوری 
(\textenglish{game theory})
ریاضی کی ايک اہم اور مروّجہ شاخ ہے جو کہ جون وان نیومین
(\textenglish{John von Neumann}) 
اورآسکر مورگن اسٹرن
(\textenglish{Oskar Morgenstern})
سےمنسوب ہے۔
گیم تھیوری عقلی فیصلہ سازوں
(\textenglish{rational decision makers})
کے درمیان تدبيری تعامل 
(\textenglish{strategic interaction})
کے ریاضیاتی ماڈل کا مطالعہ ہے اور اس کی  اطلاقات معاشيات، سياسيات، رياضياتی بائيالوجی، تجارت، انجينئرنگ، منطق
(\textenglish{logic})
اور
 کمپیوٹر سائنس کے ساتھ ساتھ کئی اور شعبوں میں بھی ہیں
۔

شطرنج، جنگ اور سیاست جیسی کھيليں انسانی تاریخ کے ہر دور ميں کھيلی گئی ہيں۔ جب بھی ايسے افراد ملتے ہيں جو متضاد خواہشات اور ترجیحات رکھتے ہوں تو ان کے درميان کھیل کھیلے جانے کے امکان ہوتے ہيں۔ مختلف کھیلوں کے تفصيلی تجزیے اور تفہيمات ایک طویل عرصے سے موجود ہيں  لیکن موجودہ دور ميں کھیلوں کے  تجزیے کا ايک رياضياتی نظریے اور ايک  رسمی مطالعہ کے طور پر سامنے آنا نسبتاء ايک حالیہ واقعہ ہے۔ کھیل کا نظریہ اصل میں اس عقلی رویے کا تجزیہ ہے جب کھيل کے شرکاء کے اعمال ايک دوسرے پر منحصر ہوں اور کھيل کے ايک شريکء کار کی حکمت عملی اس پر منحصر ہو کہ اس کے مخالفین اسی کھيل ميں کیا تدابير اختيار کرتے ہیں
\cite{Binmore,Rasmusen,Osborne}
۔
 
 کھيلوں کی عام طور پر دو واضح عمومی اشکال ہوتی ہيں۔ ايک طرح کے کھيل وہ ہيں جن ميں شريک ایجنٹوں 
 (\textenglish{agents})
 يا کھلاڑيوں کی افاديات 
 (\textenglish{utilities})
صرف امکانات  
(\textenglish{probabilities})
پر منحصر ہوتی ہيں جبکہ دوسری طرح کے وہ کھيل ہيں جن ميں شريک ایجنٹوں کی افاديات کا ان کے تدبيری اعمال
(\textenglish{strategic actions})
پر انحصار ہوتا ہيں۔ مثال کے طور پر جوا خانے
(\textenglish{casino})
ميں کھيلا جانے والا روليٹ
 (\textenglish{roulette})
ايک امکانی کھيل ہے جبکہ شطرنج ايک تدبيری کھيل ہے۔

 گیم تھیوری کی ابتدائی تاريح ايسی کھیلوں کی تحقیق سے متعلق ہے جن ميں تدبير دانوں 
 (\textenglish{strategic players})
 کی افاديّات کا مجموعہ صفر ہوتا ہے۔ يہ ايسے کھيل ہيں جن میں ایک کھلاڑی کی افاديت ميں حصول کھيل ميں شريک دوسرے کھلاڑيوں کے مجموعی نقصانات کے برابر ہوتا ہے اور ان کو صفر- جمع 
 (\textenglish{zero-sum})
 قسم کے کھیل کہا جاتا ہے۔ تاہم ايک نہ- صفر- جمع
  (\textenglish{non-zero-sum})
 کھيل ميں ایک کھلاڑی کی افاديت ميں حصول دوسرے کھلاڑيوں کے مجموعی نقصانات سے مختلف ہوتی ہے۔
 
 موجودہ دور ميں گیم تھیوری کو وسیع پیمانے پر مختلف رویّوں کے باہمی تعلقات کی  منطقانہ تشريح حاصل کرنے کيلئے لاگو کی جاتا ہے۔ يہاں تک کہ اب انسانوں، جانوروں اور کمپیوٹروں میں استدلالی فیصلے کرنے کی سمجھ بوجھ کے حصول ميں مدد کرنے والی سائنس کے طور پر اسے ایک وسيع معنوں والی اصطلاح کے طور پر بھی جانا جاتا ہے۔

 سوال یہ ہے کہ اردو ميں گیم تھیوری کا صحيح ترجمہ کيا نظریہ کھیل ہو يا کہ نظریۂ تدبير؟ گیم تھیوری کے استعمال سے جوا خانے ميں کھيلے جانے والی روليٹ 
 (\textenglish{roulette})
  ميں جيتنے کے امکانات کا تعيّن تو کيا جا سکتا ہے ليکن گيم تھيوری کے زيادہ تر معاملات کا مقصد کھيل ميں شريک  تدبير دانوں کی مختلف حکمت عملیوں 
  (\textenglish{strategies})
  میں سے صحیح حکمت عملی کا تعيّن کرنا ہوتا ہے۔ اس نقطۂ نظر سے  شايد يہ درست دکھائی ديتا ہے کہ اردو ميں  گیم تھیوری کا ترجمہ نظریۂ تدبير ہونا چاہيے نہ کہ نظریۂ کھیل۔ تاہم اس مضمون کے کچھ قارئين کو اس سے اختلاف ہو سکتا ہے۔
\section{پس منظر}
کچھ عرصہ قبل ایک دوست نے يونيورسٹی آف ايڈيليڈ ميں منعقد ايک چھوٹی سی محفل ميں يہ کہا کہ تخليق کا کام اس زبان ميں ہی بہترين ہو سکتا ہے جس ميں آپ کا دماغ سوچتا ہے اور باالخصوص وہ زبان جس ميں آپ کو خواب آتے ہيں۔ گفتگو کے پس منظر کو سامنے رکھتے ہوئے وہ يہ کہنا چاہ رہے تھے کہ ايک ايسا تخليقی عمل، جس کی نوعيت سائينسی ہو يا پھر غير سائينسی، اور جس کا بنيادی اظہار انگريزی زبان ميں ہو، اس ميں غير انگريزی پسء منظر کے لوگوں کيلئے ايسی شراکت يا حصہ داری کرنا جس کی حيثيت کو عمومی طور پر قابلء يقين اور ٹھوس سمجھا جا سکے، نہ صرف بہت مشکل ہے بلکہ اس شراکت کی حدود کے خدوخال ان لوگوں کے غير انگريزی پس منظر سے متعيّن ہوتے ہيں۔

انہوں نے معروف 
فلسفی لڈوگ وٹگنسٹائين
(\textenglish{Ludwig Wittgenstein})
کی اس بات کی طرف حوالہ ديا کہ انسانی ذہن کے بلند سوچ اور خاص طور پر اس کے گہرے اور پيچيدہ تصوّرات تک رسائی  کی صلاحيت اس انسان کے لفظوں، جملوں، اور زبان کے استعمال ميں مہارت سے متعيّن ہوتی ہے اور ظاہر ہے یہ مہارت اسکی مادری زبان ميں ہی اپنے بلند ترين درجے تک پہنچ سکتی ہے۔

یہ بہتر طور پر جاننے کيلئے کہ يہ نقطۂ نظر کيا ہے ميں نے ارادہ کيا کہ کيوں نہ اردو زبان ميں ايک مضمون لکھا جائے جو ميری اس تحقيق کے کچھ حصوں کا اردو ميں تعارف ہو جو کئی سال پہلے ميں نے قائداعظم يونيورسٹی اسلام آباد ميں بطور ايک تحقيقی طالب علم کی تھی۔ ميرے سامنے سوال يہ تھا کہ اس نوعيت کی سائينسی تحقيق جو ميں کرتا رہا ہوں کے اردو ميں اظہار سے کونسے بڑے چيلنجوں کا سامنا ہوتا ہے۔ اور خاص کر يہ کہ کيا ايسا کرنے سے وہ فوائد سامنے آتے ہيں جن کے بارے ميں وٹگنسٹائين نے ذکر کيا ہے۔ 

ايک اور وجہ يہ تھی کہ کئی سالوں سے مجھے يونيورسٹی آف ايڈيليڈ ميں ہر سال بين الاقوامی مقابلوں ميں پی ايچ ڈی سکالرشپس
(\textenglish{PhD scholarships})
جيتنے والے قابل ايرانی طالب علم نظر آتے رہے اور پاکستان سے آنے والے بہت کم نظر آئے۔ کئی ايرانی طالب علموں کے ڈيسکوں پر اعلی معيار کی سائينسی اور ٹيکنيکل مضامين پر لکھی ہوئی فارسی ميں کتابيں بھی دکھائی ديں جو انھوں نے اپنے یونيورسٹی کے دنوں ميں ايران ميں پڑھی تھيں۔ آبادی ميں پاکستان ايران سے بڑا ملک ہے ليکن ايسے مضامين پر اس معيار کی اردو ميں لکھی گئی کتابيں مجھے ياد نہيں کہ لاہور اور راولپنڈی کے اردو بازاروں ميں نظر آئی ہوں ۔ پاکستان جانا سالوں بعد ہوتا ہے اور شايد اس سلسلے ميں وہاں صورت حال اب مختلف ہو۔ ميری اردو لکھنے ميں زيادہ مہارت نہيں ہے ليکن بطور ايک پاکستانی اس زبان سے وابستگی اور شغف قدرتی عمل ہے۔ اس کے ساتھ ساتھ اس مضمون کا مواد کچھ پاکستانی طلبا و طالبات کيلئے باعثء دلچسپی ہو سکتا ہے۔  

\section{نيش کا توازن}
  نظریۂ تدبير ميں ايک اہم ترين نتيجہ 1951 ميں جون نیش 
  (\textenglish{John Nash})
  نے ثابت کيا
  \cite{JohnNash}
  جس کی رو سے ايجینٹوں کے ايک ايسے گروہ ، جس کے ارکان ايک تدبيری تعامل  
  (\textenglish{strategic interaction})
  ميں شريک ہوں، کيلئے ايک  یا اس سے زيادہ ايسے امکانی توازنات 
  (\textenglish{probabilistic equilibria})
  ہر صورت ميں متعيّن کيے جا سکتے ہيں اس طرح کہ ہر ايک امکانی توازن کی نوعيت يہ ہو کہ گروہ ميں شريک کسی ايک ايجنٹ کيلۓ اس امکانی توازن سے يکطرفہ انحراف
  (\textenglish{unilateral deviation})
  کا کوئی جواز اور محرک 
  (\textenglish{motivation})
  موجود نہ رہے۔

اگر دو کھلاڑيوں کے درميان کھيل اس طرح کا ہو کہ کھلاڑی
$A$
کے پاس
$m$
مختلف خالص تدابير
(\textenglish{pure strategies})
ہوں جبکہ کھلاڑی
$B$
کے پاس
$n$
مختلف خالص تدابير ہوں تو اس صورت میں اس کھيل ميں شامل ہر کھلاڑی کا افاديّتی ميٹرکس
(\textenglish{utility matrix})
کا آرڈر  $m\times\ n$  ہوتا ہے اور اسے اس طرح لکھا جا سکتا ہے
\cite{Binmore,Rasmusen,Osborne}
:
\begin{equation}\label{TwoPlayerPayoffMatrix}
A=(a_{ij})_{m\times n},\qquad B=(b_{ij})_{m\times n},\quad 1\leq i\leq m,\;1\leq j\leq n.
\end{equation}

اس ميٹرکس ميں موجود مقداريں یہ ظاہر کرتی ہيں کہ کھيل ميں شريک کھلاڑی 
$A$
کی افاديّت کھيل کے اختتام پر کيا ہو گی۔ ہر کھلاڑی کی افاديّت دونوں کھلاڑيوں کے مجموعی اعمال پر منحصر ہوتی ہے۔ اوپر ديے گئے افاديّتی ميٹرکس ميں کھلاڑی $A$ کی  خالص تدابير بائيں طرف والے کالم سے ظاہر ہوتی ہيں جبکہ کھلاڑی  $B$ کی خالص تداببير اوپر والی قطار سے ظاہر ہوتی ہيں۔ اس طرح کا ايک اور ميٹريکس کھلاڑی
$B$
کيلئے بھی لکھا جا سکتا ہے۔

اسی ایک مرحلے والے کھیل میں بھی ہم مخلوط تدابیر
(\textenglish{mixed strategies})
کے تصوّر کو بيان کر سکتے ہيں جو کہ خالص تدابير کے اوپر ہونے والی ايک امکانی تقسيم
(\textenglish{probability distribution})
سے حاصل ہوتا ہے۔ مرکب تدبيروں کا تصوّر یہ بيان کرتا ہے کہ ہر ممکنہ خالص تدبیر اختیار کرنے کا امکان کیا ہے؛ اس کے لیے کھیل کا بار بار کھیلا جانا ضروری نہیں۔ 

نيش کے توازن
(\textenglish{Nash equilibrium})
کو متعارف کرنے کيلئے ہميں تدبيری خاکے
(\textenglish{strategy profile})
کے تصوّر کی ضرورت ہے۔ ايک تدبيری خاکہ کھیل ميں شريک تمام کھلاڑيوں کی تدابير پر مشتمل ايک ايسے سيٹ
(\textenglish{set})
کا نام ہے جو کہ کھيل ميں کھیلی گئی تدابير اور اعمال کو مکمل طور پر بيان کرتا ہے۔ ايک تدبيری خاکے ميں ہر ايک کھلاڑی کی طرف سےکھيلی گئی صرف ايک تدبير شامل ہونی چاہيے اور یہ ايک مرکب تدبير بھی ہو سکتی ہے۔

مثال کے طور پر ايک ايسا  کھيل جس ميں شريک تين کھلاڑی ہوں اور اس کا تدبيری خاکہ 
$(s_{A}^{\ast},s_{B}^{\ast},s_{C}^{\ast})$ 
ہو
،
اس ميں
$s_{A}^{\ast}, s_{B}^{\ast}$
اور
$s_{C}^{\ast}$
تينوں کھلاڑيوں 
$A,  B$
اور 
$C$
کی باالترتيب تدابير ہوں تو اس تدبيری خاکے کے ساتھ اس کھيل ميں کھلاڑی
$A$
کی افاديت کو عمومی طور پر 
$P_{A}%
(s_{A}^{\ast},s_{B}^{\ast},s_{C}^{\ast})$
سے ظاہر کيا جاتا ہے۔

اگر ايک  تدبيری خاکہ $(s_{A}^{\ast},s_{B}^{\ast}\cdots s_{G}^{\ast})$   ايسے کھيل کيلئے ہو جس ميں $A,B\cdots G$ 
 کھلاڑی شريک ہوں اور اگر یہ  تدبيری خاکہ درج ذيل شرائط کو پورا کرے تو اسے ايک نيش کا توازن کہا جاتا ہے: 
\begin{equation}\label{Defining inequalities of NE} P_i(s_i^*,s_{-i}^*)\geq P_i(s_i,s_{-i}^*)\quad\forall i,\;\forall s_i\in S_i.\end{equation}
جب کہ يہ عدم مساواتيں
(\textenglish{inequalities})
ان تمام تدابير
$s_{A}\in S_{A}$ ، $s_{B}\in S_{B}$ 
اور
$s_{G}\in S_{G}$
کيلئے درست ہوں
،
جو کہ مرکب بھی ہو سکتی ہيں جہاں اور
$S_{A}$
،
$S_{B}$
اور
$S_{G}$
وغيرہ کھلاڑيوں کے تدبيری سيٹ ہيں۔ 

 اوپر دی گئی عدم مساواتوں سے ديکھا جا سکتا ہے کہ کوئی کھلاڑی اگر نيش کے توازن سے متعلق تدبيری خاکے
 (\textenglish{strategy profile})
 ،
  جو کہ
  $(s_{A}^{\ast},s_{B}^{\ast}\cdots s_{G}^{\ast})$
  ہے
  ،
 سے یک طرفہ طور پر انحراف کرے تو اس کی افاديت ميں اضافہ تو نہيں ليکن کمی ضرور ہو سکتی ہے۔ يعنی کھيل ميں شريک تمام کھلاڑيوں کو يہ محرّک ديا گيا ہے کہ وہ نيش کے توازن کے تدبيری خاکے کے ساتھ جڑے رہيں اور يک طرفہ طور پر اس سے انحرّاف سے گريز کريں۔   
\section{قیدیوں کا مخمصہ}
تدبيری کھيلوں کی ايک اہم مثال قیدیوں کا مخمصہ
(\textenglish{Prisoner's Dilemma})
ہے جس کی تفصيل کچھ يوں ہے
\cite{Rasmusen,Osborne}
۔ 
پوليس کے گشت کے دوران دو جرائم پيشہ افراد گليوں ميں منشيات فروخت کرتے ہوئے پکڑے جاتے ہيں۔ قانون ميں منشيات کی اس طرح فروخت کی سزا چھ ماہ قيد ہے۔ پوليس دونوں ملزموں کو مزيد تفتیش کيلئے تھانے لے جاتی ہے۔

تفتيش کے دوران ملزموں کے ريکارڈ کی چھان بين کرنے پر پوليس کو معلوم ہوتا ہے کہ دونوں ملزمان شايد ماضی ميں ايک بينک لوٹنے کی وادات ميں بھی ملوّث ہيں۔ ليکن پوليس کے پاس ان کے خلاف کوئی ايسے ٹھوس ثبوت موجود نہيں ہيں کہ جن کو استعمال کرتے ہوئے عدالت ميں جا کر انہيں مجرم ثابت کيا جا سکے۔ تاہم پوليس کو کافی حد تک يقين ہے کہ يہی وہ اصحاب ہيں جنہوں نے ماضی ميں بينک لوٹنے کی واردات کی تھی۔ کافی غورو فکر کے بعد پوليس ايک حکمتء عملی متعيّن کرتی ہے جس کی تفصيل اس طرح سے ہے
:

1۔ 
دونوں ملزمان، جن کے نام فرقان
 (\textenglish{F})
اور نزير
(\textenglish{N})
ہيں، کو عليحدہ عليحدہ کمروں ميں مزيد تفتيش کيلئے رکھا جائے۔

2۔
اپنے ساتھی کے خلاف گواہی
 دينے کو تدبير
 (\textenglish{D})
 کہا جائے جبکہ اس کے خلاف گواہی سے انکار کرنے کو تدبير 
  (\textenglish{C})
  کہا جائے۔

  3۔
اگر نزير اور فرقان دونوں اپنے عليحدہ عليحدہ کمروں سے ايک دوسرے کے خلاف بينک لوٹنے کے بارے ميں گواہی دينے سے انکار کر ديں يعنی ان کا تدبيری سيٹ
 (\textenglish{C, C})
 ہو
 ، جہاں بريکٹ ميں پہلا اندراج نزير کيلئے ہے جبکہ دوسرا فرقان کيلئے، تو دونوں کو چھ چھ ماہ قيد ميں گزارنے ہونگے جو کہ گليوں ميں منشيات بيچنے کی سزا ہے 
:
افادیات : نزير: 3 ، فرقان : 3 

4۔
فرقان کو عليحدگی ميں يہ بتایا جائے کہ اگر وہ نزير کے خلاف بينک لوٹنے کے بارے ميں گواہی دے
 (\textenglish{D})
،
جبکہ اسی دوران نزير اپنے عليحدہ کمرے ميں بينک لوٹنے کے بارے ميں جاری ايسی ہی تفتيش پر فرقان کے خلاف گواہی نہ دے
(\textenglish{C})
،
تو فرقان  کو نا صرف رہا کر ديا جائے گا بلکہ فرقان کو مبلغ ايک لاکھ روپے انعام ميں بھی ديے جائيں گے کہ اس نے پوليس کو ايک بدنام بينک کے ڈاکو (جس کا نام نزير ہے) کو پکڑنے ميں پوليس کی مدد کی۔ کيونکہ دورانء تفتيش نزير نے فرقان کے خلاف گواہی نہيں دی اس لئے نزير انصاف کی راہ ميں رکاوٹ ثابت ہوا ہے۔ اس لئے نزير کو بينک لوٹنے پر پانچ  سال سزا ہو گی
:
 افادیات : نزير: 0 ، فرقان: 5 

  5۔
اسی طرح نزير کو عليحدگی ميں يہ بتایا جائے کہ اگر وہ فرقان کے خلاف بينک لوٹنے کے بارے ميں گواہی دے
 (\textenglish{D})
 ،
جبکہ اسی دوران فرقان اپنے عليحدہ کمرے ميں بينک لوٹنے کے بارے ميں جاری ايسی ہی تفتيش پر نزير کے خلاف گواہی نہ دے
 (\textenglish{C})
 ،
تو نزير  کو نا صرف رہا کر ديا جائے گا بلکہ نزيرکو مبلغ ايک لاکھ روپے انعام ميں بھی ديے جائيں گے کہ اس نے پوليس کو ايک بدنام بينک کے ڈاکو (جس کا نام فرقان ہے) کو پکڑنے ميں پوليس کی مدد کی۔ کيونکہ دورانء تفتيش فرقان نے نزير کے خلاف گواہی نہيں دی اس لئے فرقان انصاف کی راہ ميں رکاوٹ ثابت ہوا ہے۔ اس لئے فرقان کو بينک لوٹنے پر پانچ  سال سزا ہو گی۔
:
 افادیات : نزير: 5 ، فرقان: 0

  6۔
اگر نزير اور فرقان دونوں اپنے عليحدہ عليحدہ کمروں سے ايک دوسرے کے خلاف بينک لوٹنے کے بارے ميں گواہی 
ديں
 (\textenglish{D, D})
 ،
تو دونوں کو دو دو سال قيد ميں گزارنے ہونگے۔
:
 افادیات : نزير: 1 ، فرقان: 1

   قیدیوں کے مخمصے کے اس تدبيری کھيل کو درج ذيل افادياتی ميٹريکس سے ظاہر کر سکتے ہيں
  :
\begin{equation}%
\begin{array}
[c]{c}%
\text{N}%
\end{array}%
\begin{array}
[c]{c}%
C\\
D
\end{array}
\overset{\overset{%
\begin{array}
[c]{c}%
\text{F}%
\end{array}
}{%
\begin{array}
[c]{cc}%
C & D
\end{array}
}}{\left(
\begin{array}
[c]{cc}%
(3,3) & (0,5)\\
(5,0) & (1,1)
\end{array}
\right)  } .\label{PDmatrix1}%
\end{equation}  
اس کھيل کی عمومی شکل اسطرح ہے کہ
:
\begin{equation}%
\begin{array}
[c]{c}%
\text{N}%
\end{array}%
\begin{array}
[c]{c}%
C\\
D
\end{array}
\overset{\overset{%
\begin{array}
[c]{c}%
\text{F}%
\end{array}
}{%
\begin{array}
[c]{cc}%
C & D
\end{array}
}}{\left(
\begin{array}
[c]{cc}%
(r,r) & (s,t)\\
(t,s) & (u,u)
\end{array}
\right)  }, \label{PDmatrix}%
\end{equation}
جبکہ درج ذيل مطلوبات
\begin{align}
s<u<r<t, \label{PDgen}%
\end{align}
درست ہوں۔

دونوں ملزمان خالص تدابير کھيلتے ہيں تو فرقان کی تدبير یا تو نزير کے خلاف گواہی ہو سکتی ہے يا پھر گواہی سے انکار۔ اسی طرح نزير کی تدبير یا تو فرقان کے خلاف گواہی ہو سکتی ہے يا پھر گواہی سے انکار۔ اور اس طرح افاديتی ميٹريکس
(\ref{PDmatrix1})
کے مطابق کھلاڑيوں کی افاديات ، جو کہ دونوں کی تدابير پر منحصر ہيں، کيلئے نيش کی عدم مساواتوں
(\ref{Defining inequalities of NE})
کو تدبيری جوڑے
($D$ , $D$)
 کيلئےاس طرح لکھا جا سکتا ہے
:
\begin{align}
P_{N}(D,D)-P_{N}(C,D) \leq 0,\nonumber\\
P_{F}(D,D)-P_{F}(D,C) \leq 0,
\label{DD_NE}%
\end{align}
جس ميں دی گئی ايک بريکٹ ميں پہلا اندراج کھلاڑی
$A$
کی تدبير ہے جبکہ دوسرا اندراج کھلاڑی
$B$
کی تدبير ہے۔ ميٹريکس
(\ref{PDmatrix})
کو استعمال کرتے ہوئے ان عدم مساواتوں سے ہميں
\begin{align}
(u-s)\leq0, \label{PDgen1}%
\end{align}
حاصل ہوتا ہے جو کہ مطلوبات
(\ref{PDgen})
کی روشنی ميں درست ہے۔

عدم مساواتوں
(\ref{DD_NE})
کے مطابق دونوں کھلاڑيوں کيلئے  بہتر يہی ہے کہ وہ ايک دوسرے کے خلاف گواہی ديں اور اس طرح
($D$ , $D$)
کا تدبيری جوڑا ايک نيش کے توازن کے طور پر سامنے آتا ہے۔

قیدیوں کے مخمصے ميں دلچسپ امر يہ ہے کہ باوجود اس کے کہ دونوں قيدی خاموشی کی تدابير
($C$ , $C$)
اختيار کر کے صرف چھ چھ ماہ قيد ميں گزار کر آزاد ہو سکتے ہيں، دونوں استدلالی طور پر یہی بہتر تدبير سمجھتے ہيں کہ ايک دوسرے کے خلاف گواہی دی جائے اور نتيجہ کے طور پر دونوں دو دو سال قيد ميں گزاريں۔
\section{نظريۂ تدبير اور سياست}
 نظريۂ تدبير کی اہم اطلاقات ميں معاشيات اور سياسيات شامل ہيں۔ دنيا کے چند معتبر ترين تدبير دانوں ميں اسرائيل کے پروفيسر رابرٹ آومن
(\textenglish{Robert Aumann})
کا نام شامل ہے جنہيں تھامس شيلنگ
(\textenglish{Thomas Schelling})
کے ساتھ سال  2005  ميں اکنامک سائينسز
(\textenglish{economic sciences})
ميں نوبل ميموريل پرائيز
(\textenglish{Nobel Memorial Prize in Economic Sciences})
سے نوازا گيا
-

پروفيسر آومن  اسرائيل ميں دائيں بازو کے سياسی گروپ ، جس کا نام 
"مضبوط اسرائیل کيلئے پروفیسرز"
(\textenglish{Professors for a Strong Israel})
  ہے، کے رکن ہیں۔ 2005 میں پروفيسر آومن نے اسرائيل کی غزہ ميں عدم مداخلتی تحريک کی مخالفت کی اور یہ دعوی کيا  کہ ايسا کرنا گش کٹيف
(\textenglish{Gush Katif})
کے آباد کاروں کے خلاف ایک جرم اور اسرائیل کی سلامتی کے لئے سنگین خطرہ ہے۔

پروفيسر آومن نظريۂ تدبير ميں بلیک میلر کا متناقضہ
(\textenglish{Blackmailer Paradox})
   کی طرف حوالہ ديتے ہوئے  کہتے ہیں کہ اسرائيل کی طرف سے عربوں کو زمین دينا تدبيری طور پر ايک بےوقوفانہ عمل ہے جس کی توجيح رياضياتی نظريۂ تدبير کے لٹريچر ميں  بلیک میلر کے متناقضہ کی تفصيل ميں موجود ہے
\cite{RobertAumann}
۔

       ان کے خيال ميں عرب رياستيں ایک ڈھٹائی والا  مطالبہ کر کے اسرائيل کو اس بات پر مجبور کرتی ہيں کہ وہ ان کے بليک ميل ميں آ جائے۔ اور اس عمل کی وجہ عرب رياستوں کا يہ تاثر ہے کہ اگر اسرائيل غير لچکدارانہ رويہ رکھتا ہے تو اسے گفت وشنيد کے کمرے سے خالی ہاتھ جانا ہو گا۔ 
\section{تدبير کا کوانٹم نظریہ}
کوانٹم گیم تھیوری 
(\textenglish{quantum game theory})
یا تدبير کا کوانٹم نظریہ، تحقیق کا وہ میدان ہے جس میں کھيلوں کے مروّجہ يا کلاسيکی نظریۂ کھیل کو ايسی سمت کی طرف بڑھايا جاتا ہے جوکہ روايتی طور پر کوانٹم کا علاقہ 
 (\textenglish{quantum domain})
ہے۔ تحقیق کا يہ شعبہ 1999 میں وجود میں آيا اور اب یہ کوانٹم انفارميشن اور کمپيوٹيشن 
(\textenglish{quantum information and computation}) 
کے تحقيقی ميدان
\cite{NielsenChuang}
کا حصہ ہے۔ اس تحقيقی ميدان کا بنيادی مقصد کوانٹم کمپيوٹر بنانا ہے جو کہ مستقبل ميں کلاسيکی کمپيوٹر 
(\textenglish{classical computer}) 
کے مقابلے ميں انتہائی زيادہ طاقتور ہوگا۔ رياضی کے خاص مسائل اور کوانٹم نظامات
(\textenglish{quantum systems}) 
کی سموليشن 
(\textenglish{simulation}) 
کيلئے کوانٹم کمپيوٹر کی غير معمولی طاقت کو نظرياتی طبيعات اور رياضياتی اصولوں کی پرکھ سے ثابت کيا جا چکا ہے۔ دنيا کی کئی اہم تجربہ گاہوں ميں کوانٹم کمپيوٹر بنانے کی کوششيں زور و شور سے جاری ہيں اور اسکے پروٹوٹائيپس 
(\textenglish{prototypes})
تيار کئے جا چکے ہيں۔
\subsection{ ڈيوڈ مايئر اور کوانٹم نظريۂ تدبير کی ابتدا }
اگرچہ نظريۂ کوانٹم
(\textenglish{quantum theory})
ميں ايسی صورت حال کی نشاندہی پہلے بھی متعدد بار کی گئی جس کا تجزيہ نظريۂ تدبير سے ہو سکتا ہے ليکن شايد اس ميدان ميں کام کرنے والوں کی اکثريت کے مطابق  تدبير کے کوانٹم نظریے کی باقاعدہ ابتدا سال 1999 ميں ہوئی جب يونيورسٹی آف کيليفورنيا سان ڈياگو
(\textenglish{San Diego})
 کے رياضی کے پروفيسر ڈيوڈ مايئر
(\textenglish{David Meyer})
نے طبيعات کے معتبر ترين سمجھے جانے والے جريدے فزيکل ريوّيو ليٹرز
(\textenglish{Physical Review Letters})
ميں ايک پرچہ
\cite{DavidMeyer}
شائع کيا۔

اس پرچے ميں انھوں نے اپنا یہ موقّف بيان کيا کہ کوانٹم انفارميشن اور کمپيوٹيشن
(\textenglish{quantum information and computation})
ميں ایک اوريکل مسئلہ
(\textenglish{oracle problem})
کے لئےتجویز کردہ کوانٹم الگورتھم
(\textenglish{quantum algorithm})
کو ایک ايسے تدبيری عمل کی طرح بھی سمجھا جا سکتا ہے کہ جس ميں شريک دو ایجنٹوں  کی انفرادی افادیات
(\textenglish{individual utilities})
کا مجموعہ تو صفر ہو ليکن ایک ایجنٹ کوانٹم تدابير پر عمل کر سکتا ہو جبکہ کھيل ميں شريک دوسرا ایجنٹ کلاسیکی تدابير پر عمل درآمد کرنے تک محدود ہو۔ اسطرح اگر کوانٹم الگورتھمز اپنی فطرت ميں تدبيری کھيل ہيں تو رياضياتی نظريۂ تدبير کو نئے کوانٹم الگورتھموں کی دریافت کيلئے جاری کوششوں ميں ايک نئے اوزار کے طور پر استعمال کيا جا سکتا ہے۔

کوانٹم تدبير کا نظریہ ايسے اسٹریٹجک یا باتدبير ایجنٹوں کے درمیان تعامل کا مطالعہ کرتا ہے جن کی رسائی کوانٹم انٹينگلمنٹ
(\textenglish{quantum entanglement})
اور کوانٹم سپرپوزيشن
(\textenglish{quantum superposition})
کے اہم وسائل تک بھی ہو۔ تدبير کے کوانٹم نظریہ ميں دو یا دو سے زیادہ ایجنٹوں کی طرف سے ایک کوانٹم نظام کو تدبیری طريقوں سے ترکيب کیا جاتا  ہے اور اس عمل ميں مقامی یونیٹری ٹرانسفارميشنز
(\textenglish{local unitary transformations})
اور کوانٹم  پیمائش
(\textenglish{quantum measurement})
کا اطلاق کيا جاتا ہے۔ ایجنٹوں کی انفرادی افادیات ان کے تدبیری  اقدامات (حکمت عملی)، جو کہ مقامی یونیٹری ٹرانسفارميشنز پر مشتمل ہوتی ہے، کے بعد کوانٹم سسٹم پہ کی جانے والی آخری پیمائش کے نتائج سے حاصل کی جاتی ہيں۔
\subsection{آئيزرٹ، ولکنز، اور ليونسٹائين اور  دو قیدیوں کے مخمصے کی کوانٹم شکل }
  ڈيوڈ مايئر کے
اس اہم نتيجہ کو بنياد بنا کر اسی سال يعنی 1999 ميں ہی آئيزرٹ، ولکنز، اور ليونسٹائين
(\textenglish{Eisert, Wilkens, and Lewenstein})
نے جرمنی کی پوٹس ڈيم
(\textenglish{Potsdam})
اور ہين اووّر
(\textenglish{Hannover})
يونيورسٹيوں سے فزيکل ريوّيو ليٹرز ميں ہی ايک پرچہ
\cite{EWL}
شائع کيا جس ميں انھوں نے دو قیدیوں کے مخمصے کے تدبيری کھيل
،
جس کی تفصيل اوپر بيان کی گئی ہے
،
کی ايک کوانٹم ميکانکی شکل کی تجويز کی۔ ان دو اہم پرچوں کو عموماء کوانٹم نظريۂ تدبير کی ابتدا سمجھا جاتا ہے۔ 

آئيزرٹ اور انکے ساتھيوں نے يہ دريافت کيا کہ اگر قیدیوں کے مخمصے کو اس طرح کھيلا جائے کہ دونوں کھلاڑی کی رسائی کوانٹم دنيا کے اہم وسيلے يعنی کوانٹم انٹينگلمنٹ
(\textenglish{quantum entanglement})
تک بھی ہو، اور کھلاڑيوں کی تدابير ان  کی مقامی یونیٹری ٹرانسفارميشنز
(\textenglish{local unitary transformations})
پر مشتمل ہوں تو اس تدبيری کھيل کا ايک نہايت دلچسپ نتيجہ برآمد ہوتا ہے۔ وہ يہ کہ ايک ايسا نيا نيش کا توازن اس کھيل کے حل کے طور پر سامنے آتا ہے جس ميں دونوں کھلاڑیوں (قيديوں) کی انفرادی افاديات (3 ، 3) ہوتی ہيں جو کہ کھيل کی کلاسيکی شکل ميں نا ممکن ہے چاہے کھلاڑی مرکب تدبيريں کھيلنے کے بھی مجاز ہوں ۔ ليکن اس تدبيری کھيل ميں نيش کا يہ نيا توازن دو خالص یا امکانی کلاسيکل تدابير
(\textenglish{pure or mixed classical strategies})
پر نہيں بلکہ دو خاص قسم کے مقامی یونیٹری ٹرانسفارميشنز پر مشتمل ہوتا ہے۔

آئيزرٹ اور ساتھيوں کی دو کھلاڑيوں کے درميان کھيلے گئے کوانٹم کھيل کے تعميری عمل ميں ايک کيوبٹ
(\textenglish{qubit})
کی ھلبرٹ سپيس
(\textenglish{Hilbert space})
ميں کيفيت کو دو اساسی ويکٹرز
(\textenglish{basis vectors})
سے تفويص کيا جاتا ہے جو کہ
$\left|
S_{1}\right\rangle $
اور
$\left|  S_{2}\right\rangle $
ہوتے ہيں۔ دو کيوبٹس
(\textenglish{qubits})
کی مخصوص مجموئی کيفيات
(\textenglish{states})
کو دو ايسی ھلبرٹ سپيسس
(\textenglish{Hilbert spaces})
$\mathcal{H}_{A}$
اور
$\mathcal{H}_{B}$
ميں ظاہر کيا جا سکتا ہے جس ميں ہر ايک کی ڈائیمينشن
(\textenglish{dimension})
دو عدد ہو۔ 

اس طرح کھيل کی کيفيت کو ايک ايسے ويکٹر سے بيان کيا جا سکتا ہے جو کہ ٹينسر پروڈکٹ سپيس
(\textenglish{tensor-product space})
$\mathcal{H}%
_{A}\otimes\mathcal{H}_{B}$
ميں ہو اور اس کا احاطہ اساسی ويکٹرز
$\left|  S_{1}S_{1}\right\rangle
,
\left|  S_{1}S_{2}\right\rangle
,
\left|  S_{2}S_{1}\right\rangle $
اور
$\left|  S_{2}S_{2}\right\rangle $
کريں۔
کھيل کی ابتدائی کيفيت
$\left|  \psi_{ini}\right\rangle =\hat{J}\left|  S_{1}%
S_{1}\right\rangle $
ہوتی ہے جبکہ
$\hat{J}$
ايک ايسا يونيٹری آپريٹر ہے جس کا علم دونوں کھلاڑيوں کو ہے۔ آئيزرٹ اور ساتھيوں نے يونيٹری آپريٹر
$\hat{J}$
جو کہ کھيل ميں انٹينگلمنٹ کی ايک متعيّن مقدار
$\gamma$
کو پيدا کرتا ہے کی تعريف يوں کی کہ
:
\begin{equation}
\hat{J}=\exp\left\{  i\gamma S_{2}\otimes
S_{2}/2\right\}.۔
\end{equation}

فرض کريں کہ فرقان
(\textenglish{F})
اور نزير
(\textenglish{N})
کی تدابير باالترتيب یونيٹری آپريشنز
$\hat{U}_{F}$
اور
$\hat{U}_{N}$
ہيں جن کو ايک تدبيری سپيس
(\textenglish{strategic space})
سے چنا گيا ہے جسے ہم
\textenglish{S}
سے ظاہر کرتے ہيں۔ کھلاڑيوں کے تدبيری عوامل کے بعد کھيل کی کيفيت
$\hat{U}%
_{F}\otimes\hat{U}_{N}\hat{J}\left|  S_{1}S_{1}\right\rangle $
ميں تبديل ہو جاتی ہے۔ 

کھيل کا آخری مرحلہ ايک کوانٹم پيمائش کے عمل پر مشتمل ہوتا ہے جس ميں کھيل کی تبديل شدہ کيفيت پر پہلے ايک معکوس يونيٹری آپريٹر
(\textenglish{reverse unitary operator})
کا اطلاق کيا جاتا ہے جسے
$\hat{J}^{\dagger}$
سے ظاہر کرتے ہيں، اور پھر اسے سٹرن گيرلاخ
(\textenglish{Stern-Gerlach})
کی قسم کے پيمائشی آلات کے ايک جوڑے
(\textenglish{pair})
سے گزارا جاتا ہے۔ انکشاف يا ڈيٹيکشن
(\textenglish{detection})
سے پہلے کھيل کی آخری کيفيت
(\textenglish{final state})
کو
$\left|  \psi
_{fin}\right\rangle =\hat{J}^{\dagger}\hat{U}_{F}\otimes\hat{U}_{N}\hat
{J}\left|  S_{1}S_{1}\right\rangle $
سے ظاہر کيا جاتا ہے۔

دونوں کھلاڑيوں کی متوقّع ادائيگيوں
(\textenglish{expected utilities})
کو کھيل کی آخری کيفيت
$\left|  \psi_{fin}\right\rangle $
کو ٹينسر پروڈکٹ سپيس
(\textenglish{tensor-product space})
،
جو کہ
$\mathcal{H}_{A}%
\otimes\mathcal{H}_{B}$
ہے
،
کے اساسی ويکٹرز کے اوپر پروجيکشنز
(\textenglish{projections})
کرنے سے اس طرح حاصل کيا جاتا ہے کہ ان پروجيکشنز کے اوزان
(\textenglish{weights})
وہ مستقل مقداريں ہوں جو کہ کھيل کے ميٹريکس
(\textenglish{matrix})
ميں موجود ہيں۔ مثال کے طور پر ادائيگی ميٹريکس
(\ref{PDmatrix})
کو استعمال کرتے ہوئے نزير
(\textenglish{N})
کی متوقّع ادائيگی
(\textenglish{payoff})
کو اسطرح لکھا جاتا ہے
:
\begin{equation}
P_{N}=r\left|  \left\langle S_{1}S_{1}\mid\psi_{fin}\right\rangle \right|
^{2}+s\left|  \left\langle S_{1}S_{2}\mid\psi_{fin}\right\rangle \right|
^{2}+t\left|  \left\langle S_{2}S_{1}\mid\psi_{fin}\right\rangle \right|
^{2}+u\left|  \left\langle S_{2}S_{2}\mid\psi_{fin}\right\rangle \right|  ^{2},
\label{Furqan's payoff}%
\end{equation}
اور یوں فرقان
(\textenglish{F})
کی متوقّع ادائيگی کو ہم اس ٹرانفارميشن
(\textenglish{transformation})
سےحاصل کر سکتے ہيں
:
$s\rightleftarrows t$
،
يعنی کہ 
:
\begin{equation}
P_{F}=s\left|  \left\langle S_{1}S_{1}\mid\psi_{fin}\right\rangle \right|
^{2}+t\left|  \left\langle S_{1}S_{2}\mid\psi_{fin}\right\rangle \right|
^{2}+s\left|  \left\langle S_{2}S_{1}\mid\psi_{fin}\right\rangle \right|
^{2}+u\left|  \left\langle S_{2}S_{2}\mid\psi_{fin}\right\rangle \right|  ^{2}.
\label{Nazir's payoff}%
\end{equation}
آئيزرٹ اور ساتھيوں نے دو کھلاڑيوں کے درميان کھيل کی اپنی اصلی تجويز کردہ کوانٹم شکل ميں دونوں کھلاڑيوں کی تدابير کو درج ذيل يونيٹری آپريٹرز کے سيٹ تک محدود کيا
:
\begin{equation}
U(\theta,\phi)=\left(
\begin{tabular}
[c]{ll}%
$e^{i\phi}\cos(\theta/2)$ & $\sin(\theta/2)$\\
$\text{-}\sin(\theta/2)$ & $e^{-i\phi}\cos(\theta/2)$%
\end{tabular}
\right),  \label{Eisert's unitary operators}%
\end{equation}
جس ميں
$\theta\in\lbrack0,\pi]$
اور
$\phi\in\lbrack0,\pi/2]$
ہو۔ اس طرح اگر
$\gamma\in\lbrack0,\pi/2]$
ہو
،
جو کہ کھيل ميں موجود انٹينگلمنٹ کی مقدار کو ظاہر کرتا ہے۔

اب اگر يونيٹری آپريٹرز کا یہ جوڑا
\begin{align}
(\hat{U}_{A}^{\ast},\hat{U}_{B}^{\ast}) \label{NE_QG}
\end{align}
 ايک نيش کا توازن ہو تو اس صورت ميں
\begin{align}
\Pi_{A}(\hat{U}_{A}^{\ast},\hat{U}_{B}^{\ast})-\Pi_{A}(\hat{U}_{A},\hat{U}%
_{B}^{\ast})\geq0,\nonumber\\
\Pi_{B}(\hat{U}_{A}^{\ast},\hat{U}_{B}^{\ast
})-\Pi_{B}(\hat{U}_{A}^{\ast},\hat{U}_{B})\geq0,
\label{NashInequalities}%
\end{align}
ہوگا یعنی کہ دونوں کھلاڑيوں ميں سے ہر ايک کيلئے تدبيری جوڑے
(\ref{NE_QG})
سے یکطرفہ انحرّاف کرنے سے کوئی فائدہ تو نہيں ليکن نقصان ضرور ہو سکتا۔

آئيزرٹ اور ساتھيوں نے کھلاڑيوں کی تدابير کو جس يونيٹری آپريٹرز کے سيٹ
(\ref{Eisert's unitary operators})
تک محدود کيا اس کی توجيہہ پر بعد ميں دوسرے تحقيق دانوں نے سوالات اٹھائے ليکن اس کی تفصيل پر گفتگو اس مضمون کی دسترس
(\textenglish{scope})
سے باہر ہے۔ 

دلچسپ بات يہ ہے کہ جب انٹينگلمنٹ کی پيمائش
$\gamma=0$ 
ہو تو کوانٹم کھيل وہ کلاسيکل شکل اختيار کر ليتا ہے جس ميں دونوں کھلاڑی مرکب تدبيريں کھيلنے کے مجاز ہوں۔ آئيزرٹ اور ساتھيوں نے یہ معلوم کيا کہ کھيل ميں سب سے زيادہ انٹينگلمنٹ يعنی  
$\gamma=\pi/2$
کی صورت ميں ايک یکتا
(\textenglish{unique})
پيريٹو آپٹيمل
(\textenglish{Pareto optimal})
توازن ايک يونيٹری آپريٹرز کے جوڑے
$\hat{Q}\otimes\hat
{Q}$ 
پر مشتمل ہوتا ہے اس طرح کہ
$\hat{Q}\sim\hat{U}(0,\pi/2)$
ہو
۔
    پيريٹو آپٹيمل حل کسی کھيل کا ايسا نتيجہ ہوتا جسے کم از کم کسی ايک شريک کھلاڑی کی افاديت کو گھٹائے بغير کم نہيں کيا جا سکتا
    \cite{Rasmusen}
    ۔

يعنی قیدیوں کے مخمصے کے تدبيری کھيل کو اگر کوانٹم دنيا کے اہم وسيلے يعنی کوانٹم انٹينگلمنٹ کو استعمال کر کے کھيلا جائے تو دونوں کھلاڑی (قيدی)  ايک دوسرے کے خلاف گواہی دينے کے استدلالی نتيجے سے نہ صرف باہر نکل سکتے ہيں بلکہ ايک ايسا پيريٹو آپٹيمل
(\textenglish{Pareto-optimal})
نتيجہ
:
\begin{align}
(\hat{Q},\hat{Q}),
\end{align}
بھی حاصل کر سکتے ہيں کہ جس ميں دونوں کھلاڑیوں (قيديوں) کوصرف چھ چھ ماہ قيد ميں گزارنے ہونگے۔ يہ سزا ان کے گليوں ميں منشيات بيچنے کيلئے ہے اوراسکی گواہ خود پوليس ہے۔ يعنی 
\begin{align}
P_{N}(\hat{Q},\hat{Q})  &  =3=P_{F}(\hat{Q},\hat{Q}).
\end{align}
اس تدبيری کھيل کی کلاسيکی شکل ميں، جس ميں دونوں کھلاڑيوں کی رسائی مرکب تدبيروں تک بھی ہو، ايسا نتیجہ بطور ايک نيش کے توازن کے حاصل کرنا ناممکن ہے۔

اب تک کے دريافت شدھ کوانٹم الگورتھموں
\cite{NielsenChuang}
کی فہرست زيادہ طويل نہيں اور یہ وہ مسائل ہيں جن کو حل کرنے کے دوران کوانٹم کمپيوٹر کی کلاسيکی کمپيوٹر پر واضح سبقت ثابت کی جا سکتی ہے۔ تدبير کے کوانٹم نظریہ کو نئے کوانٹم الگورتھموں کو دریافت کرنے کيلئے جاری موجودہ کوششوں میں پہلے سے دستیاب اور زير استعمال ریاضیاتی آلات ميں تدبير کے کلاسیکی  نظریہ کی اضافت
(\textenglish{addition})
کے طور پر بھی ديکھا جا سکتا ہے۔

یہ اس وجہ سے اہميت کا حامل ہے کہ تدبير کے کلاسیکی نظریہ ميں ایجنٹوں کے درميان پيچيدہ باہمی معاملات کے حل کيلۓ طاقتور اور گہرے ریاضیاتی تصورات موجود ہيں اور جيسا کہ ڈيوڈ مائير نے دکھايا کہ کوانٹم الگورتھموں کو کلاسیکی اور کوانٹم باتدبير ایجنٹوں کے درميان پيچيدہ باہمی عوامل کے طور پر بھی سمجھا جا سکتا ہے
\cite{MeyerQGQA}
۔
\section{ارتقائی طور پر مستحکم تدابير}
سال
1993
ميں برطانيہ کی شيفيلڈ يونيورسٹی
(\textenglish{University of Sheffield})
ميں طالب علمی کے دوران ميری ملاقات وہاں کے رياضی کے ڈيپارٹمنٹ ميں اس وقت کے ايک پوسٹ ڈاکٹرل فيلو ڈاکٹر مارک بروم
(\textenglish{Dr Mark Broom})
سےہوئی تھی۔ ان کی پی ايچ  ڈی آکسفورڈ يونيورسٹی سے تھی اور وہ  جون مينارڈ سمتھ
(\textenglish{John Maynard Smith}) 
کے ارتقائی عوامل پر رياضياتی تصورات کے استعمال پر تحقيق کر رہے تھے۔ ان کی تحقيق کا موضوع ارتقائی عوامل کے رياضياتی تصوّرات کو استعمال کرتے ہوئے پرندوں کےگھونسلے بنانے کے خاص رويّوں کی تشريح کرنا تھا
\cite{MBroom}
 ۔
  یہ موضوع مجھے بہت حيرت انگيز لگا کہ کيسے ايک رياضياتی تصوّر کو ايسے مقصد کيلئے بھی استعمال کيا جا سکتا ہے۔ مارک بروم آج کل لندن کی سٹی يونيورسٹی
(\textenglish{City University})
ميں ارتقائی رياضيات کے پروفيسر ہيں
۔

1970
کی دھائی ميں کلاسيکی نظریۂ تدبير کو بڑی کاميابی کے ساتھ جانوروں کی دنيا ميں ہونے والے ارتقائی عوامل کی سمجھ بوجھ حاصل کرنے کيلئے استعمال کيا گيا۔ اس سلسلے ميں جون مينارڈ سمتھ
،
جن کا تعلق برطانيہ کی سسّکس يونيورسٹی
(\textenglish{University of Sussex})
سے رہا، کا نام پوری دنيا ميں جانا جاتا ہے۔ خاص کر ارتقائی بائيالوجی ميں ارتقائی طور پر مستحکم تدابير
(\textenglish{Evolutionarily Stable Strategies})
(ای ایس ایس)  کے رياضياتی تصور کو مشہور اور مروّج کرنے ميں جون مينارڈ سمتھ کا نام  مرکزی حيثيت رکھتا ہے
\cite{MaynardSmith}
۔

ایک ارتقائی طور پر مستحکم تدبير ایک ایسی تدبير ہے جسے کسی ماحول میں موجود ايجينٹوں يا کھلاڑيوں کی ايک آبادی کی طرف سے کھيلنے کيلئے اگر ايجينٹوں کی طرف سے اگر منظور کر ليا جائے تو اس پر کوئی اور متبادل اور متغير تدبير حملہ آور نہیں ہو سکتی جو کہ اسی آبادی ميں موجود ايک ايسے چھوٹے گروپ کی طرف سے کھيلی جائے جس کا اس آبادی ميں تناسب بہت کم ہو۔ 

اسطرح ايک آبادی جو کہ ايک ارتقائی طور پر مستحکم تدبير کھيل رہی ہو وہ ايک چھوٹے گروہ کی طرف سے کھیلی جانے والی ايک مختلف اور متغير تدبير کی کاميابی سے مزاحمت کر سکتی ہے۔ ارتقائی طور پر مستحکم تدبير کے تصور کی جڑيں نظريۂ تدبير اور جنس کے تناسب
(\textenglish{sex ratio})
کے موضوع پر پہلے سے کی گئی تحقيق سے جا ملتی ہيں اور اس سلسلے ميں معروف رياضی دان رونالڈ فشر
(\textenglish{Ronald Fisher})
کا نام باالخصوص جانا جاتا ہے جو جديد شماريات
(\textenglish{statistics})
کے بانی بھی سمجھے جاتے ہيں
\cite{RFisher}
 ۔
 قاری کيلئے شايد يہ باعثء دلچسپی ہو کہ رونالڈ فشر سال
 1957 
 اور
 1962
 کے دوران يونيورسٹی آف ايڈيليڈ سے وابستہ رہے اور ان کی باقيات ايڈيليڈ شہر ميں واقع سينٹ پيٹرز کيتھڈرل
(\textenglish{St. Peter's Cathedral})
کے اندر ايک بينچ کے پاس مدفون ہيں۔

مينارڈ سمتھ نے ايک ايسی بڑی آبادی کا رياضياتی مطالعہ کيا جس کے ارکان یا کھلاڑی بے ترتيبانہ طور پر ليکن جوڑوں کی شکل ميں اسطرح ملائے جاتے ہيں کہ ہر ملاپ پر دونوں ارکان کے درميان ايک تدبيری کھيل کھیلا جاتا ہے جسے ايک دوہرے ميٹريکس
(\textenglish{bimatrix})
سے ظاہر کر سکتے ہيں ۔ 

کھلاڑی گمنام
(\textenglish{anonymous})
ہوتے ہيں اور اسطرح کھلاڑيوں کے کسی جوڑے کے درميان کھيلے گئےتدبيری کھيل کو ايک تشاکلی دوہرے ميٹريکس
(\textenglish{symmetric bi-matrix})
سے ظاہر کر سکتے ہيں ۔ کھلاڑی اپنی تدابير اور ادائيگيوں کے حوالے سے ايک دوسرے سے مماثل ہوتے ہيں۔ 

دوہرے ميٹريکس کے تشاکلی ہونے کا مطلب ہے کہ فرض کيا کہ دونوں کھلاڑيوں کا تدبيری سيٹ
 $S$
 ہے۔  کھلاڑی
A
کی تدبير
$S_{1}\in S$
ہے جبکہ کھلاڑی
B
کی تدبير
$S_{2}\in S$
ہے۔ تو کھلاڑی
A
کی ادائیگی اس ادائيگی کے برابر ہو گی جو کہ کھلاڑی
B
کو کی جائيگی جب کھلاڑی
A
کی تدبير
$S_{2}$
ہو اور کھلاڑی
B
کی تدبير
$S_{1}$
ہو۔

نظريۂ تدبير ميں ايک  تشاکلی دوہرے ميٹريکس کو
 $G=(M,M^{T})$
 سے ظاہر کرتے ہيں جس ميں
 $M$
 کھلاڑی 
 A
 کی ادائيگی کا ميٹريکس ہے جبکہ
 $M^{T}$
 ،
 جو کہ اس کا ٹرانسپوز
 (\textenglish{transpose})
ہے
،
 وہ کھلاڑی
B
کی ادائيگی کا ميٹريکس ہے۔ کھلاڑيوں کے جوڑوں کے ايک تشاکلی مقابلے
(\textenglish{symmetric pair-wise contest})
ميں اگر ايک
$x$
کھلاڑی کی ادائيگی
،
جبکہ اس کا مقابل
$y$
کھلاڑی ہو
،
کو
$P(x,y)$
سے ظاہر کرتے ہيں۔

ايسے تشاکلی مقابلے
(\textenglish{symmetric contest})
ميں دو کھلاڑيوں کے مابين تدابير کے تبادلے سے دونوں کھلاڑيوں کی ادائيگیوں ميں تبادلہ ہوجاتا ہے۔ اسطرح سے ايک کھلاڑی کی ادائيگی اس کی تدبير سے متعين ہوتی ہے نہ کہ اس کی شناخت سے اور کھلاڑيوں کی ادائيگيوں کو ظاہر کرنے کيلئے  اندراجء ذيلی
(\textenglish{subscript})
کی ضرورت نہيں رہتی۔

 ايک تدبير
$x$
کو ايک ارتقائی طور پر مستحکم تدبير کہا جائے گا اگر ہر دوسری تدبير
$y$
،
جبکہ
$y\neq x$
ہو
،
کيلئے درج ذيل عدم مساوات
:
\begin{equation}
P[x,(1-\epsilon)x+\epsilon y]>P[y,(1-\epsilon)x+\epsilon y] \label{ESSDefIneq},%
\end{equation}
ہر حسب ضرورت حد تک چھوٹے
$\epsilon>0$
کيلئے درست ہو۔ اس عدم مساوات کے بائيں طرف تدبير
$x$
کی وہ ادائيگی ہے جب اس تدبير کو ايک دوسری تدبير
$(1-\epsilon)x+\epsilon y$
کے مقابل کھيلا جائے اور
$\epsilon\in\left[  0,\epsilon
_{0}\right)  $
ہو۔
اس صورت ميں جب
$\epsilon>\epsilon_{0}$
ہو تو يہ کہا جاتا ہے کہ ايک متغير تدبير
(\textenglish{mutant strategy})
حملہ آور ہے۔ مقدار
$\epsilon_{0}$
کو حملہ آور یا چڑھائی کرنے والی تدبير کے خلاف رکاوٹ
(\textenglish{invasion barrier})
کہا جاتا ہے۔

اس تصوّر کو ايسے بھی بيان کر سکتے ہيں کہ ايک تدبير
$x$
ارتقائی طور پر مستحکم  ہوتی ہے اگر ہر
$y$
متغير تدبير
(\textenglish{mutant strategy})
کی طرف سے ہونے والی چڑھائی کيخلاف ايک مثبت رکاوٹ موجود ہو۔

حملہ آور متغيّر تدبير کی چڑھائی کے خلاف  اس حالت تک مثبت رکاوٹ موجود رہتی ہے کہ آبادی ميں متغير تدبير
$y$
کھيلنے والے ارکان کا تناسب چڑھائی کے خلاف اس مثبت رکاوٹ سے کم ہو جائے۔ ايسی صورت ميں تدبير
$(1-\epsilon)x+\epsilon y$
کے خلاف تدبير
$x$
کی متوقع ادائيگی تدبير
$y$
کی متوقع ادائيگی سے زيادہ ہوتی ہے۔

ارتقائی طور پر مستحکم تدبير
$x$
کے حصول کيلئے مندرجہ بالا شرائط کو درج ذيل دو مطلوبات
\cite{MaynardSmith}
کے طور پر لکھا جا سکتا ہے
:
\begin{align}
1.\text{ \ \ \ }P(x,x)  &  >P(y,x),\nonumber\\
2.\text{ If\ }P(x,x)  &  =P(y,x)\ \text{then}\ P(x,y)>P(y,y). \label{DefESS}%
\end{align}
اسطرح ارتقائی طور مستحکم تدبير ايک ايسے  نيش کے متشاکل توازن
(\textenglish{symmetric Nash equilibrium})
کے طور پر سامنے آتی ہے جس ميں چھوٹے تغيرات
(\textenglish{small mutations})
کے خلاف توازن
(\textenglish{stability})
کی خاصيت ہوتی ہے
\cite{Weibull}
۔

مطلوبات
(\ref{DefESS}) 
کی پہلی شرط يہ دکھاتی ہے کہ اگر
$x$
ايک ارتقائی طور مستحکم تدبير ہے تو تدبيری جوڑا
$(x,x)$
اس دوہرے ميٹريکس والے کھيل
$G=(M,M^{T})$
کيلئے نيش کا ايک توازن ہو گا۔ تاہم اس کا متضاد درست نہيں ہو گا۔ یعنی اگر تدبيری جوڑا
$(x,x)$
 نيش کا ايک توازن ہو تو
 $x$
ايک ارتقائی طور مستحکم تدبير صرف اس صورت ميں ہو گی کہ تدبير 
$x$
مطلوبات
(\ref{DefESS})
میں دی گئی دوسری شرط کو بھی پورا کرے۔

ارتقائی طور مستحکم تدابير یا ای ايس ايس
(\textenglish{ESS})
کے تصوّر نے ارتقائی نظريۂ تدبير
(\textenglish{evolutionary game theory})
کو مروّج کرنے ميں سب سے بڑے محرّک کا کام کيا۔ موجودہ دور ميں ای ايس ايس کے تصوّر کو ايک آبادی کے متعامل افراد
(\textenglish{interacting individuals})
کی مابين ارتقائی حرکيات
(\textenglish{evolutionary dynamics})
کو سمجھنے کيلئے بطور ايک مرکزی ماڈل کے جانا جاتا ہے۔
ای ايس ايس کا تصوّر اس سوال کو اٹھاتا اور اسکا جواب ديتا ہے کہ ايک آبادی کی کون سی ايسی متوازن  کيفيات
(\textenglish{stable states})
نموندار ہونگی جب اس آبادی ميں چنّاو کا ايسا طريقۂ کار عمل پيرا ہو جو بہتر تدبيروں کا طرف دار ہو؟ يہاں متوازن کيفيات وہ ہيں جو تغيّرات
(\textenglish{mutations})
کی وجہ سے ہونے والے انتشار
(\textenglish{perturbations})
کے خلاف مزاحمتی اہليت رکھتی ہيں۔

ای ايس ايس کے نظريہ کا متحرّک چارلس ڈارون
(\textenglish{Charles Darwin})
کا قدرتی چنّاو
(\textenglish{natural selection})
کا طريقۂ کار ہے جسے ايک ايلگورتھم
(\textenglish{algorithm})
کی صورت ميں ترکيب ديا جاسکتا ہے جسے ريپليکيٹر ڈائينيمک
(\textenglish{replicator dynamic})
کہتے ہيں
\cite{Hofbauer}
۔
ريپليکيٹر ڈائينيمک کی اہم خاصيت بے ترتيبانہ طور پر تغيرّات کے عمل سے گزرتے ہوئے ريپليکيٹرز
(\textenglish{replicators})
ميں چناو کا دہراو یا اعادہ
(\textenglish{iterations})
ہے۔

يہ ڈائينيمک ايک رياضياتی بيان ہے جو يہ کہتا ہے کہ ايک آبادی ميں ان کھلاڑيوں کا تناسب وقت کے ساتھ بڑھے گا جو بہتر تدبيريں کھيلتے ہيں۔ ايک آبادی جس ميں چناو کا بنيادی طريقۂ کار ريپليکيٹر ڈائينيمک ہو اس ميں ارتقائی طور مستحکم تدابير بطور ان تدبيروں کے سامنے آتی ہيں جن ميں تغيرّات کی طرف سے پيدا کردہ انتشار کے خلاف اپنا توازن برقرار رکھنے کی اہليت ہوتی ہے۔ مناسب مفروضوں میں ایسی تدبیریں ریپلیکیٹر حرکیات کے بعض مستحکم ساکن نقاط سے متعلق ہوتی ہیں؛ تاہم ہر ساکن نقطہ لازماً ارتقائی طور پر مستحکم نہیں ہوتا۔ اس پہلو کو حرکیاتی نظام کے ساکن نقاط
(\textenglish{rest points})
اور ان کے استحکام کی الگ جانچ سے واضح کیا جاسکتا ہے
\cite{Cressman}
۔    

مزيد تجزيے پر معلوم ہوتا ہے کہ ايک ارتقائی طور مستحکم تدبير کا تصوّر اصل ميں متشاکل نيش توازنات
(\textenglish{symmetric Nash equilibria})
کے سيٹ پہ ايک باريک بينی
(\textenglish{refinement})
ہے۔ يعنی ہر ارتقائی طور مستحکم تدبير ايک متشاکل نيش کا توازن تو ہوتی ہے ليکن ہر متشاکل نيش کا توازن ايک ارتقائی طور مستحکم تدبير نہيں ھوتا
\cite{Cressman}
۔  
\section{شاعرء مشرق اور تدبيرو تقدير}
شاعری ميں ميرا علم اور شوق بہت محدود ہے ليکن تدبير سے متعلق شاعرء مشرق علامہ اقبال کے کچھ اشعار نظر سے گزرے۔ مثلاء مختلف جگہوں پر فرماتے ہيں کہ 
:
\begin{center}
آزمودہ فتنہ ہے اک اور بھی گردوں کے پاس
\\
 سامنے   تقدير   کے  رسوائئ  تدبير    ديکھ
\end{center}
\begin{center}
ذرّہ   ذرّہ  دہر  کا  زندانئ  تقدير ہے
\\
پردۂ مجبوری و بے چارگی تدبير ہے
\end{center}
\begin{center}
آثار تو کچھ کچھ نظر آتے ہيں کہ آخر
\\
تدبير کو تقدير کے شاطر نے کيا مات
\end{center}

ان اشعار سے لگتا یہی  ہے کہ شايد شاعرء مشرق تقدير کو تدبير پر مقدّم سمجھتے تھے۔ ارتقائی طور پر مستحکم تدبيروں کی تفصيل جاننے پر يہ  معلوم ہوتا ہے کہ کيسے ارتقائی عوامل خود ہی ايک اعلی اور اپنے ماحول سے زيادہ مطابق تدبير کو وقت گزرنے کے ساتھ مستحکم کرنے ميں مدد کرتے ہيں۔ اور جب يہ استحکام حاصل کر ليتی ہے تو کوئی اور چھوٹے گروپ کی طرف سے کھيلی گئی ايک مختلف تدبير اس ارتقائی طور پر مستحکم تدبير کے خلاف موثّر ثابت نہيں ہو سکتی۔ اس سے يہ سوال ابھرتا ہے کہ کيا وہ تدبير جس نے اپنے آپ کو ارتقائی طور پر مستحکم کر ليا ہے وہ اقبال کے اوپر درج شعروں کی تقدير تو ہی نہيں؟  
\section{ارتقائی طور پر مستحکم تدابير کی کوانٹم حيئيت}
اتفاق سے سال 2000 کے شروع  ميں تدبيرکے کوانٹم نظريے پر شائع ہونے والے چند شروع کے اور اہم پرچے
\cite{DavidMeyer,EWL,MeyerQGQA}
ميری نگاہ سے گزرے تو مجھے مارک بروم کے شيفيلڈ يونيورسٹی ميں تحقيقی کام اور جون مينارڈ سمتھ کے ارتقائی طور پر مستحکم تدابير کے رياضياتی تصوّر کی ياد آئی۔ ميری رائے ميں اگر تدبير کا کوانٹم نظريہ تجويز کيا جا چکا ہے تو اس کا مطلب يہ ہوا کہ ارتقائی طور پر مستحکم تدابير، جن کی ریاضياتی بائيالوجی ميں بہت اہم اور وسيع اطلاقات ہيں،  کے تصور کی کوانٹم کی سمت ميں توسيع ممکن ہونی چاہيے۔ ميں نے مارک بروم سے ای ميل پر رابطہ کيا اور ای ايس ايس کے موضع پر لکھے گئے جون مينارڈ سمتھ اور مارک بروم کے پرچے پڑھنا شروع کئے۔ اگلے دو سالوں کے دوران ڈاکٹر عبدالحميد طور صاحب کے زيرنگرانی ميں نے ارتقائی طور پر مستحکم تدابير کے تصور کی تشريح  تدبير کے نئے تجويز کردہ کوانٹم نظريے ميں تلاش شروع کی اور اس موضع پر اپنا کام
\cite{IqbalToor2001a,IqbalToor2001b,IqbalToor2002a,IqbalToor2002c,IqbalToor2004a}
قائداعظم يونيورسٹی اسلام آباد کے اليکٹرانکس ڈيپارٹمنٹ سے شائع کيا۔

مثال کے طور پر اب ہم ايک ايجنٹس
(\textenglish{agents})
يا کھلاڑيوں کی ايسی آبادی کو ديکھتے ہيں جس ميں ايک ای ايس ايس نے اپنے آپ کو مستحکم کر ليا ہے۔ ہم يہ جاننا چاہتے ہيں کہ اس صورت ميں کيا ہو گا کہ اگر اس آبادی ميں کھلاڑيوں کا ايک ايسا چھوٹا متغيّر تدابير کھيلنے والا گروہ نمودار ہو جائے جو کہ کوانٹم تدابير بھی کھيل سکتا ہو۔ کيا ایسا گروہ مستحکم کلاسيکی ای ايس ايس پر حملہ آور ہو سکے گا؟ کيا اس چھوٹے گروہ کا حملہ کامياب رہے گا اور کيا اس صورت ميں ايسی ای ايس ايس اپنے آپ کو مستحکم کر لے گی جو اپنی نوعيت ميں کوانٹم مکينيکی ہے؟

کوانٹم ای ايس ايس کے مستحکم ہونے کے بعد ايک اور چھوٹا گروہ نموندار ہوتا ہے جو کہ ايک اور مختلف متغير کوانٹم تدبير کھيلنے کی اہليت رکھتا ہے۔ کيا يہ گروہ اپنی کوانٹم تدبير کو استعمال کرتے ہوئے پہلے سے مستحکم کوانٹم ای ايس ايس پر حملہ آور ہو سکے گا
\cite{IqbalToor2001a}
۔ 

ايک ايسی آبادی کو سامنے رکھتے ہوئے جس ميں جوڑا جوڑا متشاکل نوعيت کے مقابلے
(\textenglish{symmetric pairwise contests})
جاری ہوں ہم درج ذيل ميں اوپر بيان کردہ سوالوں کا تجزيہ کرتے ہيں۔ اس سلسلے ميں ہم قیدیوں کے مخمصے کی تدبيری کھيل کو ديکھتے ہيں جسے 
آئيزرٹ
،
ولکنز، اور ليوئنسٹائين کی سکيم استعمال کرتے ہوئے کوانٹم طريقے سے کھيلا گيا ہے۔   

جيسا کہ اوپر ہم نے ديکھا کہ قیدیوں کے مخمصے کے تدبيری کھيل ميں ہر قيدی کے پاس 
$C$
يا
$D$
کی خالص تدبيريں ہوتی ہيں۔ ايک آبادی جس ميں يہ کھيل ہر ہونے والے جوڑا جوڑا مقابلوں ميں کھيلا جا رہا ہو اس ميں کون سی تدبيريں ارتقائی طور پر مستحکم رہيں گی؟ یہ جاننے کيلئے ہم قیدیوں کے مخمصے کے ميٹرکس
(\ref{PDmatrix1})
اور ای ايس ايس ہونے کی مطلوبات
(\ref{DefESS})
سے يہ ديکھتے ہيں کہ
\begin{equation}
1= P(D,D)  >P(C,D) = 0,
\end{equation}
یعنی
(\ref{DefESS})
 ميں ای ايس ايس کی پہلی شرط پورا ہونے پر اس آبادی ميں خالص تدبير
 $D$
 بطور ای ايس ايس سامنے آئے گی۔ 
 
 آئيزرٹ ، ولکنز، اور ليوئنسٹائين کی سکيم ميں بھی
 کھلاڑيوں کی ادائيگی کا ميٹريکس
 (\ref{PDmatrix1})
ہے ليکن کھلاڑيوں کی ادائيگيوں کی مساواتوں ميں موجود امکانات
(\textenglish{probabilities})
کوانٹم ميکانيکی ہيں۔ کھيل کی اس کوانٹم شکل کے ايک آبادی ميں ارتقائی نتائج کو ديکھنے کيلئے ہم اس صورت حال کو ديکھتے ہيں جس ميں آبادی کےارکان کے درميان جوڑا جوڑا مقابلے جاری ہوں اور ہر ايسے مقابلے ميں قیدیوں کے مخمصے کا کھيل کھيلا جاتا ہو۔ خالص تدبير
$D$
اپنے آپ کو بطور ايک ای ايس ايس مستحکم کر چکی ہو۔ اب اس آبادی ميں ايک چھوٹا گروہ نموندار ہوتا ہے جو کہ کوانٹم تدبير
$\hat{U}(\theta,\phi)$
کھيلنے کی اہليت رکھتا ہے جو کہ مساوات
(\ref{Eisert's unitary operators})
ميں بيان کی گئی ہے۔

يہ جاننے کيلئے کہ کيا کوانٹم تدبير
$\hat{U}(\theta,\phi)$
مستحکم ای ايس ايس
$D$
پر حملہ آور ہو سکے گی يا نہيں ہم مساواتوں
(\ref{Furqan's payoff})
،
(\ref{Nazir's payoff})
،
اور
(\ref{PDmatrix1})
،
(\ref{PDmatrix})
میں ديے گئے ميٹريسيز
(\textenglish{matrices})
کی روشنی ميں درج ذيل ادائيگيوں کو ديکھتے ہيں
:
\begin{align}
P(D,D)  &  =1,\nonumber\\
P(D,\hat{U}(\theta,\phi))  &  =5\cos^{2}(\phi)\cos^{2}(\theta/2)+\sin
^{2}(\theta/2),\nonumber\\
P(\hat{U}(\theta,\phi),D)  &  =5\sin^{2}(\phi)\cos^{2}(\theta/2)+\sin
^{2}(\theta/2),\nonumber\\
P(\hat{U}(\theta,\phi),\hat{U}(\theta,\phi))  &  =3\left|  \cos(2\phi)\cos
^{2}(\theta/2)\right|  ^{2}+5\cos^{2}(\theta/2)\sin^{2}(\theta/2)\left|
\sin(\phi)-\cos(\phi)\right|  ^{2}+\nonumber\\
&  \left|  \sin(2\phi)\cos^{2}(\theta/2)+\sin^{2}(\theta/2)\right|  ^{2}.%
\end{align}
ان مساواتوں کو ای ايس ايس کی تعريفی عدم مساواتوں
(\ref{DefESS})
کے تناظر ميں ديکھنے پر معلوم ہوتا ہے کہ اگر
$\phi<\arcsin(1/\sqrt{5})$
ہو تو اس صورت ميں
$P(D,D)>P(\hat{U}(\theta,\phi),D)$
ہو گا۔ اور اگر
$P(D,D)=P(\hat{U}(\theta,\phi),D)$
ہو تو اس صورت ميں
$P(D,\hat{U}(\theta,\phi))>P(\hat
{U}(\theta,\phi),\hat{U}(\theta,\phi))$
ہو گا۔

اسلئے تدبير
$D$
ايک ای ايس ايس کے طورپر مستحکم رہے گی جب تک
$\phi<\arcsin(1/\sqrt{5})$
ہو وگرنہ تدبير
$\hat{U}(\theta,\phi)$
اس قابل ہو جائے گی کہ وہ مستحکم ای ايس ايس
$D$
پر حملہ آور ہو سکے۔

اس تجزيے ميں حملہ آور متغير تدبير کھيلنے والے کھلاڑيوں کے پاس ايک زيادہ ثمر آور کوانٹم ميکانيکی تدبير ہے جس کے نتيجے ميں اگر
$\phi>\arcsin(1/\sqrt{5})$
ہو تو مستحکم ای ايس ايس 
$D$
جو اپنی نوعيت ميں کلاسیکی ہے
،
پر حملہ کامیاب رہتا ہے۔

ليکن يہ مشاہدہ ہميں ايک اور صورت حال کی طرف لے جاتا ہے کہ جس ميں ايک متغير کوانٹم تدبير
$\hat{Q}\sim\hat{U}(0,\pi/2)$
کھيلنے والا ايک چھوٹا گروہ ايک کامياب حملہ کرتے ہوئے اپنے آپ کو بطور ايک ای ايس ايس مستحکم کر ليتی ہے۔ کيا اب ايک اور چھوٹا گروہ کوانٹم متغير تدبير
$\hat{U}(\theta,\phi)$
کھيلتے ہوئے پہلے سے مستحکم کوانٹم ای ايس ايس
$\hat{Q}\sim\hat{U}(0,\pi/2)$
پر کامیابی کے ساتھ حملہ آور ہو سکے گا؟

اس سوال کا جواب دينے کيلئے ہم يہ قاری کو ياد دلاتے ہیں کہ قیديوں کے دو عارظہ کھيلنے کی کوانٹم سکيم ميں دونوں کھلاڑیوں کی طرف سے کھيلی جانے والی کوانٹم تدبير
 $\hat{Q}\sim\hat{U}(0,\pi/2)$
 بطور ايک يکتا نيش کے توازن کے سامنے آتی ہے۔ اب
 $\hat{Q}$
کے مقابلے ميں متغيّر کوانٹم تدبير 
$\hat{U}(\theta,\phi)$
 کھيلنے والے ايک چھوٹے گروہ کی کيا صورت حال ہو گی؟ اسکا جواب دينے کيلئے ہم درج ذيل ادائيگياں معلوم کرتے ہيں 
:
\begin{align}
P(\hat{Q},\hat{Q})  &  =3,\nonumber\\
P(\hat{U}(\theta,\phi),\hat{Q})  &  =[3-2\cos^{2}(\phi)]\cos^{2}%
(\theta/2),\nonumber\\
P(\hat{Q},\hat{U}(\theta,\phi))  &  =[3-2\cos^{2}(\phi)]\cos^{2}%
(\theta/2)+5\sin^{2}(\theta/2).
\end{align}

 اسطرح عدم مساوات
$P(\hat{Q},\hat{Q})>P(\hat{U}(\theta,\phi),\hat{Q})$
ان تمام
$\theta\in\lbrack0,\pi]$
اور
$\phi\in\lbrack0,\pi/2]$
کيلئے درست رہتی ہے سوائے اس صورت کے جب
$\theta=0$
اور
$\phi=\pi/2$
ہوں اور يہ وہ صورت حال ہے جب متغير کوانٹم تدبير
$\hat{U}(\theta,\phi)$
اور
$\hat{Q}$
ايک ہو جائيں اور اسے ہم نظر انداز کر سکتے ہيں۔ اسطرح مطلوبات
(\ref{DefESS})
کو سامنے رکھتے ہوئے ان کی پہلی شرط
$\hat{Q}$
کے بطور ايک ای ايس ايس کے مستحکم ہونے کيلئے کافی ہے۔ مطلوبات کی دوسری شرط کے مطابق
$P(\hat{Q},\hat{Q})=P(\hat{U}(\theta,\phi),\hat{Q})$
اور جسکا مطلب يہ ہوا کہ
$\theta=0$
اور
$\phi=\pi/2$
۔
ايک دفعہ پھر يہ وہ صورت حال ہے کہ جس میں متغير تدبير وہی ہے جيسے کہ
$\hat{Q}$
اور اس صورت حال کو ہم نظر انداز کر سکتے ہيں۔ اور اسطرح کوانٹم تدبير
$\hat{U}(\theta,\phi)$
ايک کوانٹم ای ايس ايس
$\hat{Q}\sim
\hat{U}(0,\pi/2)$
پر کاميابی سے حملہ آور نہيں ہو سکتی۔ 

کوانٹم تدابير اور ارتقائی استحکام کے موضوع پر کوچی يونيورسٹی آف ٹيکنالوجی
(\textenglish{Kochi University of Technology})
ميں ٹاکسو چی
اون
(\textenglish{Taksu Cheon})
کے ساتھ ميرا لکھا ہوا کتاب کا ايک باب حوالہ
\cite{AzharChapter}
پر دستياب ہے۔
 \section{کوانٹم نظريۂ تدبير پر کام کرنے والے پاکستانی نام}

قائداعظم يونيورسٹی اسلام آباد کے اليکٹرانکس ڈيپارٹمنٹ سے پی ايچ ڈی کے دوران احمد نواز صاحب سے تعارف ہوا اور انہوں نے بھی ڈاکٹر عبدالحميد طور صاحب کے زير نگرانی اپنی پی ايچ ڈی کيلئے کوانٹم نظريۂ تدبير کا موضوع چنا اور اس وجہ سے ہمارے درميان دلچسپ اور مفيد گفتگو کی  کئی نشستيں قائداعظم یونيورسٹی ميں جاری رہيں۔ کوانٹم نظريۂ تدبير پر احمد نواز صاحب کے پرچے اپنی چھپنے سے پہلے کی شکل ميں حوالہ
\cite{AhmadNawazArxiv}
پر دستياب ہيں۔     

احمد نواز صاحب آجکل اسلام آباد ميں تحقيق اور ترقی کے ايک ادارے سے وابستہ ہيں۔ ان کے بعد قائداعظم یونيورسٹی کے طبيعات کے ڈيپارٹمنٹ سے ڈاکٹر خالد خان صاحب کے زير نگرانی محمد رمضان صاحب اور سلمان خان صاحب نے بھی اپنی پی ايچ ڈيز کيلۓ کوانٹم نظريۂ تدبير کے موضوع کا انتخاب کيا۔ محمد رمضان صاحب بھی آجکل اسلام آباد ميں ايک تحقيق اور ترقی کے ايک ادارے سے وابستہ ہيں اور ان کے پرچے اپنی چھپنے سے پہلے کی شکل ميں حوالہ
\cite{MRamzanArxiv}
پر دستياب ہيں۔ سلمان خان صاحب آجکل کامسيٹس انسٹيٹيوٹ آف انفارميشن ٹيکنالوجی اسلام آباد
(\textenglish{Comsats Institute of Information Technology Islamabad})
سے وابستہ ہيں۔ ان کے پرچے اپنی چھپنے سے پہلے کی شکل ميں حوالہ
\cite{SalmanKhanArxiv}
پر دستياب ہيں۔ 

    فيصل شاہ خان صاحب کا تعلق پاکستان اور امريکہ سے ہے اور انھوں نے امريکہ کی پورٹلينڈ سٹيٹ یونيورسٹی
(\textenglish{Portland State University})
سے کوانٹم نظريۂ تدبير اورمتعلقہ موضوعات پرپی ايچ ڈی کی ڈگری حاصل کی اور آجکل ابو ذھبی کی خليفہ يونيورسٹی
(\textenglish{Khalifa University Abu Dhabi})
سے وابستہ ہيں۔ فيصل شاہ خان صاحب کے پرچے اپنی چھپنے سے پہلے کی شکل ميں حوالہ 
\cite{FaisalShahKhan}
پر دستياب ہيں۔ قائداعظم يونيورسٹی اسلام آباد سے ڈاکٹر طور صاحب کے ساتھ اور یونيورسٹی آف ايڈيليد ميں ساتھيوں کے ساتھ ميرے پرچے اپنی چھپنے سے پہلے کی شکل ميں حوالہ
\cite{AzharArxiv}
پر موجود ہيں۔

\section{توازن، استحکام اور کوانٹم تدبیر: ایک ضروری امتیاز}
نیش کا توازن بنیادی طور پر \textbf{بہترین انفرادی ردِعمل} کی شرط ہے: جب دوسرے کھلاڑی اپنی موجودہ تدابیر پر قائم رہیں تو کوئی ایک کھلاڑی تنہا اپنی تدبیر بدل کر زیادہ افادیت حاصل نہ کر سکے۔ اس کے برعکس ارتقائی استحکام میں سوال یہ ہے کہ کسی بڑی آبادی میں مستعمل تدبیر کیا کسی نایاب متبادل تدبیر کی دراندازی کو روک سکتی ہے؟ یہ دونوں سوالات قریب ضرور ہیں مگر یکساں نہیں۔ ایک توازن نیش کی شرط پوری کر سکتا ہے لیکن ارتقائی طور پر غیرمستحکم رہ سکتا ہے \cite{MaynardSmith,Weibull}۔

دو کھلاڑیوں کے متقارن کھیل میں اگر $P(x,y)$ سے مراد تدبیر $x$ اختیار کرنے والے کی وہ متوقع افادیت ہو جو اسے تدبیر $y$ کے مقابلے میں ملتی ہے، تو تدبیر $x$ ارتقائی طور پر مستحکم ہے اگر ہر مختلف متبادل $y$ کے لیے یا تو
\begin{equation}\label{ess-refined-a}
P(x,x)>P(y,x),
\end{equation}
یا برابری کی صورت میں
\begin{equation}\label{ess-refined-b}
P(x,x)=P(y,x)\quad\text{اور}\quad P(x,y)>P(y,y)
\end{equation}
ہو۔ یہ شرائط بتاتی ہیں کہ صرف ایک توازن ڈھونڈنا کافی نہیں؛ ہمیں یہ بھی دیکھنا پڑتا ہے کہ نایاب متبادل کے خلاف مقیم آبادی کی کامیابی کا تقابل کیا ہے۔

\subsection{کوانٹم کھیل میں قابلِ اجازت تدابیر}
کوانٹم کھیل بناتے وقت صرف ابتدائی کوانٹم کیفیت طے نہیں کی جاتی، بلکہ یہ بھی واضح کرنا ہوتا ہے کہ کھلاڑی کون سی تبدیلیاں استعمال کر سکتے ہیں۔ مثال کے طور پر دو پیرامیٹر والے مقامی یونیٹری اعمال کی جماعت، تمام $SU(2)$ یونیٹری اعمال کی جماعت سے چھوٹی ہے۔ کسی محدود جماعت میں موجود نیش کا توازن ضروری نہیں کہ بڑی جماعت میں بھی برقرار رہے۔ چنانچہ یہ بیان کہ کوانٹم وسائل سے قیدیوں کا مخمصہ حل ہو جاتا ہے، صرف متعلقہ کھیل کے واضح طور پر بتائے گئے اصولوں اور تدبیری حدود کے اندر سمجھنا چاہیے \cite{EWL,BenjaminHaydenComment}۔

ایک عمومی کوانٹم طریقۂ کار میں ابتدا میں دو کیوبٹ کی مشترکہ کیفیت $\rho_{\mathrm{in}}$ تیار کی جاتی ہے، کھلاڑی مقامی کوانٹم اعمال $\mathcal{E}_A$ اور $\mathcal{E}_B$ اختیار کرتے ہیں، اور آخر میں نتائج $k$ کی پیمائش $M_k$ سے ہوتی ہے۔ اگر نتیجے $k$ پر کھلاڑی $i$ کی افادیت $a_i(k)$ ہو تو
\begin{equation}\label{general-quantum-payoff}
P_i(\mathcal{E}_A,\mathcal{E}_B)=\sum_k a_i(k)\,\operatorname{Tr}\!\left[M_k(\mathcal{E}_A\otimes\mathcal{E}_B)(\rho_{\mathrm{in}})\right].
\end{equation}
اس اظہار میں $\{M_k\}$ ایک مکمل مثبت پیمائش \textenglish{(POVM)} ہے جس کے اثرات کا مجموعہ شناختی عامل کے برابر ہو۔ اس کلیے کا فائدہ یہ ہے کہ ابتدا کی کیفیت، دستیاب اعمال، اور آخری پیمائش کو الگ الگ دیکھا جا سکتا ہے؛ اس لیے کسی بھی نتیجے کو صرف انٹینگلمنٹ کی موجودگی سے منسوب کرنے سے پہلے ان تینوں اجزا کا کردار معلوم کرنا ضروری ہے۔

\subsection{مساوی افادیت والی تدابیر اور نئی تحقیق}
کوانٹم تدابیر کے ساتھ ایک اور باریک مسئلہ پیدا ہوتا ہے: ممکن ہے دو مختلف یونیٹری عوامل ہر ممکن حریف کے مقابلے میں یکساں افادیت دیں۔ ایسی صورت میں یہ طے کرنا ضروری ہے کہ انہیں حیاتیاتی یا تدبیری لحاظ سے دو جدا متبادل سمجھا جائے یا افادیت کے اعتبار سے ایک ہی تدبیر کی دو نمائندگیوں کے طور پر لیا جائے۔ اس فیصلے سے \textbf{سخت} نیش توازن اور ارتقائی استحکام کے نتائج متاثر ہو سکتے ہیں، اگرچہ عام نیش کی شرط محض افادی تقابل پر قائم رہتی ہے۔ ایک حالیہ تحقیق میں محدود آیزرٹ--ولکنز--لیونسٹائن تدبیری جماعت، مکمل $SU(2)$ جماعت اور افادی لحاظ سے مماثل تدابیر کی شناخت کے بعد بننے والی مختصر جماعت کے نتائج کے فرق کو واضح کیا گیا ہے \cite{IqbalAbbott2026}۔

یہ نکتہ کوانٹم نظریۂ تدبیر کے مطالعے میں ایک اہم اصول تجویز کرتا ہے: ہر دعوے کے ساتھ یہ بھی بتایا جائے کہ تدبیر کس چیز کو کہتے ہیں، کون سے اعمال جائز ہیں، افادیت کیسے ماپی جاتی ہے، اور مختلف ناموں یا عوامل سے ظاہر ہونے والی تدابیر کو کب ایک ہی تصور کیا جائے گا۔

\section{نتائج کی تعبیر اور تجرباتی حدود}
کوانٹم کھیلوں کی ریاضیات میں ایک تدبیر کا بہتر قرار پانا بذاتِ خود یہ ثابت نہیں کرتا کہ حقیقی معاشی، سیاسی، یا حیاتیاتی فیصلہ ساز کوانٹم انٹینگلمنٹ استعمال کر رہے ہیں۔ اس کے لیے قابلِ عمل طبعی نظام، ابتدائی کیفیت کی تیاری، قابلِ اعتماد اعمال، شور اور ماحولیاتی تعامل \textenglish{(decoherence)} کی مقدار، اور نتائج کی تجرباتی پیمائش درکار ہوتی ہے۔ اسی طرح ارتقائی استحکام کی تعبیر کے لیے آبادی میں افراد کے اختلاط، افزائش یا نقل کے طریقے اور افادی فرق سے آبادی کے تناسب میں تبدیلی کا ربط متعین کرنا ضروری ہے۔

مثلاً اگر آبادی میں تدبیر $x$ کا تناسب $z_x$ اور اس کی اوسط افادیت $f_x$ ہو، تو ایک عام ماڈل میں ریپلیکیٹر حرکیات یوں لکھی جاتی ہیں:
\begin{equation}\label{replicator-new}
\dot z_x=z_x\bigl(f_x-\bar f\bigr),\qquad \bar f=\sum_y z_y f_y.
\end{equation}
یہ ایک ممکنہ حرکی مفروضہ ہے، کوانٹم میکانیات کا لازمی نتیجہ نہیں۔ اسی لیے نیش توازن، ارتقائی استحکام اور حرکی استحکام کو الگ الگ جانچنا مناسب ہے \cite{Hofbauer,Cressman}۔

\section{قارئین کے لیے مطالعے کا نقشہ: کن سوالات سے آغاز کیا جائے؟}
اس مضمون کو سمجھنے کے لیے کوانٹم طبیعیات کے تمام پیچیدہ مباحث سے پیشگی واقفیت ضروری نہیں۔ البتہ چند بنیادی سوالات ذہن میں رکھنا مفید ہے۔ پہلا سوال یہ ہے کہ جب دو انسانوں یا دو اداروں کے مفادات ایک دوسرے پر منحصر ہوں تو ان کی عقلی تدبیر کیا ہوگی؟ دوسرا یہ کہ اگر فیصلے محض ایک مرتبہ کرنے کے بجائے بار بار آبادی میں آزمائے جائیں تو کون سی تدبیر برقرار رہے گی؟ تیسرا یہ کہ جب کھیل کے اصول میں کوانٹم کیفیات اور کوانٹم پیمائش شامل کردی جائیں تو ممکنہ نتائج کا مجموعہ کس طرح بدلتا ہے؟

یوں اس مضمون کی تین سطحیں بنتی ہیں: \textbf{تدبیری سطح}، جس میں افادیت اور بہترین ردِعمل کا حساب ہوتا ہے؛ \textbf{ارتقائی سطح}، جس میں آبادی کی ساخت اور وقت کے ساتھ تبدیلی کا مطالعہ ہوتا ہے؛ اور \textbf{طبعی سطح}، جس میں اس بات کا تعین ہوتا ہے کہ کسی کوانٹم طریقۂ کار کو حقیقی تجربہ گاہ میں کس طرح نافذ کیا جاسکتا ہے۔ ان سطحوں کو ایک دوسرے سے الگ رکھنا سائنسی احتیاط کی پہلی شرط ہے۔

\section{کوانٹم معلومات کی مختصر مگر بنیادی لغت}
\subsection{بِٹ، کیوبٹ اور پیمائش}
عام کمپیوٹر کی ایک کلاسیکی بِٹ صفر یا ایک ہوتی ہے۔ اس کے برخلاف ایک کیوبٹ کی خالص کیفیت دو بنیادی حالتوں کے خطی امتزاج سے ظاہر ہوتی ہے:
\begin{equation}\label{qubit-reader}
|\psi\rangle=\alpha|0\rangle+\beta|1\rangle,\qquad |\alpha|^2+|\beta|^2=1.
\end{equation}
یہاں $\alpha$ اور $\beta$ پیچیدہ اعداد ہوسکتے ہیں۔ اگر اس کیوبٹ کو $\{|0\rangle,|1\rangle\}$ کی بنیاد میں ناپا جائے تو صفر حاصل ہونے کا امکان $|\alpha|^2$ اور ایک حاصل ہونے کا امکان $|\beta|^2$ ہے۔ صرف اس بات سے کہ کیفیت میں دونوں اجزا شامل ہیں، یہ نتیجہ نکالنا درست نہیں کہ پیمائش سے بیک وقت دو قطعی جوابات حاصل ہوجائیں گے۔ ایک پیمائش ایک نتیجہ دیتی ہے؛ امکانات جاننے کے لیے عموماً تجربہ متعدد بار دہرایا جاتا ہے \cite{NielsenChuang}۔

\subsection{کوانٹم اشتراک (انٹینگلمنٹ) کیا ہے؟}
دو الگ کیوبٹ کی ایک مشترکہ کیفیت کی مثال یہ ہے:
\begin{equation}\label{bell-reader}
|\Phi^+\rangle=\frac{|00\rangle+|11\rangle}{\sqrt{2}}.
\end{equation}
اس کیفیت کو دو انفرادی خالص کیوبٹ کیفیات کے سادہ ٹینسر حاصلِ ضرب کے طور پر نہیں لکھا جاسکتا؛ یہی اس کے انٹینگلڈ ہونے کی نشانی ہے۔ دونوں کیوبٹ کو ایک ہی حسابی بنیاد میں ناپنے سے $00$ اور $11$ کے نتائج مساوی امکان، یعنی $1/2$، سے ملتے ہیں؛ $01$ اور $10$ کے نتائج نہیں ملتے۔ لیکن مشترکہ تعلق کی موجودگی خودبخود کسی پیغام کو روشنی سے تیز رفتار سے بھیجنے کا ذریعہ نہیں بنتی۔ نیز صرف ایک بنیاد میں مطابقت رکھنے والے نتائج انٹینگلمنٹ کا تجرباتی ثبوت نہیں: کلاسیکی طور پر باہم مربوط بِٹس بھی ایسی مطابقت دے سکتی ہیں۔ اس امتیاز کے لیے متعدد پیمائشی بنیادوں اور مناسب تجرباتی آزمائشوں کی ضرورت ہوتی ہے۔

\subsection{یونیٹری عمل، مخلوط کیفیت اور ماحولیاتی شور}
بند کوانٹم نظام میں تدبیری تبدیلی عموماً ایک یونیٹری عامل $U$ سے ظاہر کی جاتی ہے جس کی خاصیت $U^\dagger U=I$ ہے۔ مگر اصل تجربے میں محض یونیٹری عمل ہی کافی نہیں ہوتا: تیاری کی غلطیاں، غیر کامل آلات اور ماحول کے ساتھ تعامل سے کیفیت مخلوط ہوسکتی ہے۔ ایسی صورت میں کثافتی مصفوفہ $\rho$ استعمال ہوتا ہے، جس کے لیے $\rho\succeq0$ اور $\operatorname{Tr}\rho=1$ ضروری ہیں۔

اس فرق سے ایک قاری بنیادی بات سمجھ سکتا ہے: \textbf{کوانٹم تدبیر کا نام}، \textbf{کوانٹم آلے کی اصل کارروائی} اور \textbf{حاصل شدہ اعداد و شمار} تین الگ چیزیں ہیں۔ ایک مدلل تحقیق میں ان کا باہمی ربط واضح ہونا چاہیے۔

\section{قیدیوں کے مخمصے کا مکمل حساب: توازن کیوں پیدا ہوتا ہے؟}
قیدیوں کے مخمصے کا ایک معروف نمونہ ذیل کی جدول میں دیا گیا ہے۔ پہلی عددی افادیت پہلے کھلاڑی $A$ کی، اور دوسری $B$ کی ہے۔ تعاون کے لیے $C$ اور انحراف کے لیے $D$ لکھا گیا ہے:
\begin{center}
\begin{tabular}{c|cc}
 & $B:C$ & $B:D$ \\ \hline
$A:C$ & $(3,3)$ & $(0,5)$ \\
$A:D$ & $(5,0)$ & $(1,1)$
\end{tabular}
\end{center}
اس مثال میں جب $B$ تعاون کرتا ہے تو $A$ کو تعاون سے $3$ اور انحراف سے $5$ ملتے ہیں؛ اس لیے $A$ کے لیے انحراف بہتر ہے۔ جب $B$ انحراف کرتا ہے تب بھی $A$ کو تعاون سے $0$ اور انحراف سے $1$ ملتا ہے۔ اسی دلیل کا اطلاق $B$ پر بھی ہوتا ہے۔ اس بنا پر $(D,D)$ نیش کا توازن ہے اور دونوں کو $(1,1)$ ملتا ہے، اگرچہ $(C,C)$ پر $(3,3)$ دونوں کے لیے بہتر ہے۔ اس فرق کو \textbf{انفرادی محرک اور اجتماعی فائدے کا تضاد} کہا جاسکتا ہے \cite{Osborne,EWL}۔

اب فرض کریں کہ $B$، احتمال $q$ کے ساتھ تعاون کرتا ہے۔ تب $A$ کو تعاون کی متوقع افادیت
\begin{equation}\label{classical-coop}
P_A(C;q)=3q+0(1-q)=3q
\end{equation}
اور انحراف کی متوقع افادیت
\begin{equation}\label{classical-defect}
P_A(D;q)=5q+1(1-q)=1+4q
\end{equation}
ملتی ہے۔ فرق $P_A(D;q)-P_A(C;q)=1+q>0$ ہر $0\le q\le1$ پر مثبت ہے۔ لہٰذا اس مخصوص عددی کھیل میں انحراف نہ صرف خالص بلکہ تمام کلاسیکی مخلوط تدابیر کے مقابلے میں غالب ہے۔

یہ حساب اس لیے اہم ہے کہ کسی کوانٹم توسیع میں اگر ایک نیا توازن سامنے آتا ہے تو ہمیں یہ دیکھنا پڑے گا کہ \textbf{کھیل کی کون سی ریاضیاتی یا عملی شرط بدلی ہے}۔ محض افادیت کی جدول کو دوبارہ لکھ دینا کوانٹم برتری ثابت نہیں کرتا۔

\section{کوانٹم کھیل کی تعمیر: ایک قدم بہ قدم طریقۂ کار}
کوانٹم طریقۂ کار کو صرف ``انٹینگلمنٹ کے ساتھ کھیل'' کہہ دینا نامکمل ہے۔ ایک واضح ماڈل کی تیاری کے لیے کم از کم درج ذیل اجزا متعین کرنا ہوں گے:
\begin{enumerate}[label=\arabic*.]
\item \textbf{کلاسیکی بنیاد:} کھلاڑی، دستیاب فیصلے اور ہر ممکن نتیجے کی افادیت۔
\item \textbf{ابتدائی کیفیت:} مثال کے طور پر $|00\rangle$ یا اس سے تیار ہونے والی انٹینگلڈ کیفیت۔
\item \textbf{تدبیری اعمال:} ہر کھلاڑی کن یونیٹری یا عمومی کوانٹم عملیاتی نقشوں میں سے انتخاب کرسکتا ہے؟
\item \textbf{آخری پیمائش:} کون سی پیمائش انجام دی جائے گی اور نتیجے کو کس کلاسیکی جواب کے ساتھ جوڑا جائے گا؟
\item \textbf{افادیت کا قاعدہ:} ہر پیمائشی جواب سے افادیت کیسے حساب ہوگی؟
\item \textbf{تقابلی معیار:} کلاسیکی کھیل میں کون سی باہمی رسائی، قبل از وقت رابطہ، مشترکہ اتفاق اور بے ترتیب وسائل دستیاب تھے؟
\end{enumerate}
ایک مثال کے طور پر دو کیوبٹ کے خالص حالتی طریقے میں
\begin{equation}\label{protocol-reader}
|\psi_f\rangle=J^\dagger(U_A\otimes U_B)J|00\rangle
\end{equation}
لکھا جاسکتا ہے۔ حسابی بنیاد کی پیمائش پر $ij$ نتیجہ ملنے کا امکان
\begin{equation}\label{protocol-prob}
p_{ij}=|\langle ij|\psi_f\rangle|^2,\qquad \sum_{i,j\in\{0,1\}}p_{ij}=1
\end{equation}
ہے۔ اگر کھلاڑی $A$ کی افادیت $a_{ij}$ ہو تو
\begin{equation}\label{protocol-payoff}
P_A(U_A,U_B)=\sum_{i,j\in\{0,1\}}a_{ij}p_{ij}
\end{equation}
بنتی ہے۔ اس عمل کی حقیقی معنویت اسی وقت واضح ہوتی ہے جب $J$، $U_A$، $U_B$، پیمائش اور ہر افادیت کی صراحت کی جائے۔ آیزرٹ اور ساتھیوں کے نتائج انہی انتخابوں کے تحت سمجھے جاتے ہیں؛ عمومی $SU(2)$ تدابیر کی اجازت دینے پر بعض دعوے تبدیل ہوجاتے ہیں \cite{EWL,BenjaminHaydenComment,EisertReply}۔

\subsection{کوانٹم فائدہ اور منصفانہ کلاسیکی مقابلہ}
ایک اہم سوال یہ ہے کہ اگر کوانٹم کھیل کسی کلاسیکی کھیل سے بہتر نتیجہ دیتا ہے تو \textbf{بہتر کس معنی میں}؟ بعض اوقات کلاسیکی ماڈل میں کھلاڑی صرف آزادانہ بے ترتیب انتخاب کرسکتے ہیں، جبکہ کوانٹم ماڈل میں پہلے سے مشترکہ حالت فراہم کی جاتی ہے۔ ایک منصفانہ موازنہ پوچھتا ہے کہ اگر کلاسیکی کھلاڑیوں کو بھی پہلے سے مشترکہ بے ترتیب علامت یا محدود رابطے کی سہولت ملے تو کیا نتیجہ برقرار رہتا ہے؟ اگر نہیں، تو فرق کو فوراً انٹینگلمنٹ کا غیر معمولی فائدہ کہنا مناسب نہیں۔

کوانٹم کھیل کے لیے \textbf{غیر مقامی تعلقات}، \textbf{انٹینگلمنٹ}، \textbf{مشترکہ کلاسیکی ربط} اور \textbf{تدبیری توازن} الگ الگ تصورات ہیں۔ ان کے درمیان تعلقات کا مطالعہ نہایت اہم ہے، مگر ایک کا وجود دوسرے کو لازماً ثابت نہیں کرتا \cite{KhanReview2018}۔

\section{ارتقائی استحکام کی حسابی مثال: باز اور فاختہ}
ارتقائی نظریۂ تدبیر کی اہمیت اس وقت زیادہ نمایاں ہوتی ہے جب تدبیری ``عقل مندی'' کے بجائے آبادی میں کامیابی کو دیکھا جائے۔ ایک سادہ \textenglish{Hawk--Dove} کھیل تصور کریں جس میں وسائل کی قدر $V>0$ اور لڑائی کی ممکنہ لاگت $C>V$ ہو۔ باز (H) جارحانہ مقابلہ کرتا ہے جبکہ فاختہ (D) تصادم سے بچنے کی کوشش کرتی ہے۔ روایتی مفروضوں کے تحت پہلی تدبیر اختیار کرنے والے کی افادیت یہ ہوگی:
\begin{center}
\begin{tabular}{c|cc}
 & $H$ & $D$ \\ \hline
$H$ & $(V-C)/2$ & $V$ \\
$D$ & $0$ & $V/2$
\end{tabular}
\end{center}
اگر آبادی میں بازوں کا تناسب $x$ ہو تو باز کی اوسط کامیابی
\begin{equation}\label{hawk}
f_H(x)=x\frac{V-C}{2}+(1-x)V
\end{equation}
اور فاختہ کی اوسط کامیابی
\begin{equation}\label{dove}
f_D(x)=x\cdot0+(1-x)\frac V2
\end{equation}
ہے۔ ان کا فرق
\begin{equation}\label{hawkdiff}
f_H(x)-f_D(x)=\frac{V-Cx}{2}
\end{equation}
نکلتا ہے۔ اس لیے اندرونی توازن
\begin{equation}\label{hawkeq}
x^*=\frac VC \qquad (0<x^*<1)
\end{equation}
ہے۔ اگر $V=2$ اور $C=4$ رکھا جائے تو $x^*=1/2$ ملتا ہے۔ اس کا مطلب یہ نہیں کہ ہر جانور بیک وقت نصف باز اور نصف فاختہ ہے۔ ایک تعبیر یہ ہے کہ آبادی میں نصف افراد باز کی اور نصف فاختہ کی تدبیر اختیار کرتے ہیں؛ دوسری تعبیر میں ہر فرد تصادفی طور پر دونوں رویوں میں سے کسی ایک کو چنتا ہے، بشرطیکہ ماڈل کے مفروضے اس تعبیر کی اجازت دیں۔

\subsection{ریپلیکیٹر مساوات سے استحکام کی تصدیق}
دو تدبیروں کے لیے ریپلیکیٹر مساوات خاص طور پر سادہ ہے:
\begin{equation}\label{hawkrep}
\dot x=x(1-x)[f_H(x)-f_D(x)]=\frac12x(1-x)(V-Cx).
\end{equation}
اگر $0<x<x^*$ ہو تو $\dot x>0$، یعنی بازوں کا تناسب بڑھتا ہے؛ اور اگر $x^*<x<1$ ہو تو $\dot x<0$، یعنی یہ تناسب گھٹتا ہے۔ اس طرح $x^*$ اندرونی آبادی کے لیے مقامی طور پر متقارب توازن ہے۔ اس مثال سے \textbf{نیش توازن، ارتقائی استحکام اور آبادی کی حرکیات} کے رشتے کا عملی تصور ملتا ہے؛ تاہم ان تصورات کی عمومی مساوات فرض کرنا درست نہیں \cite{MaynardSmith,Hofbauer,Weibull}۔

کوانٹم ارتقائی کھیل میں بھی اگر ہمیں افادی مصفوفہ یا افادی تابع حاصل ہوجائے تو اسی قسم کی حرکی جانچ کی جاسکتی ہے۔ البتہ اس سے پہلے یہ طے کرنا لازمی ہے کہ آبادی میں قابلِ وراثت تدبیر سے کیا مراد ہے: ایک مقامی عمل، اس عمل کی افادی جماعت، یا کوئی تجرباتی رویہ؟

\section{شور، نامکمل انٹینگلمنٹ اور نتائج کی مضبوطی}
عملی کوانٹم تجربات میں کامل کیفیات ہمیشہ دستیاب نہیں ہوتیں۔ فرض کریں ابتدائی حالت $\rho_{\mathrm{ideal}}$ ہے مگر شور کے باعث ہم ایک سادہ نمونے میں
\begin{equation}\label{wernerlike}
\rho_v=v\rho_{\mathrm{ideal}}+(1-v)\frac{I_4}{4},\qquad 0\le v\le1
\end{equation}
استعمال کرتے ہیں۔ $v=1$ میں مثالی حالت اور $v=0$ میں مکمل مخلوط حالت حاصل ہوتی ہے۔ یہ محض ایک مخصوص مخلوط شور کا نمونہ ہے؛ تمام حقیقی آلودگیوں کی نمائندگی نہیں کرتا۔ اگر باقی طریقۂ کار اور پیمائشی افادیت کا عامل $G_i$ مستقل رکھا جائے تو افادیت کا تعلق
\begin{equation}\label{noiseaffine}
P_i(v)=\operatorname{Tr}(G_i\rho_v)
=vP_i(1)+(1-v)P_i(0)
\end{equation}
خطی ہوگا۔ لیکن \textbf{نیش توازن} کا شور کے ساتھ خطی بدلنا ضروری نہیں، کیونکہ بہترین ردِعمل کے تقابل میں متعدد تدابیر شریک ہوتی ہیں۔ کسی خاص $v$ پر توازن پیدا، ختم یا غیر منفرد ہوسکتا ہے۔ اسی لیے صرف مثالی $v=1$ کے نتائج کو مکمل عملی دعویٰ نہیں سمجھنا چاہیے۔

\subsection{ایک قابلِ آزمائش تحقیقی سوال}
فرض کریں کسی محدود تدبیری جماعت میں ایک تدبیر $Q$ نیش توازن ہے۔ اس دعوے کو مضبوط کرنے کے لیے یہ سوالات مفید ہیں: کیا $Q$ مکمل دستیاب عملیاتی جماعت میں بھی بہترین ردِعمل ہے؟ کیا یہ توازن شور کی چھوٹی تبدیلی کے بعد برقرار رہتا ہے؟ کیا اس کا افادی فائدہ کلاسیکی مشترکہ تصادف سے بھی حاصل کیا جاسکتا ہے؟ کیا تجرباتی طور پر اتنی درست پیمائش دستیاب ہے کہ دو تدابیر کے افادی فرق کو قابلِ اعتماد طور پر الگ کیا جاسکے؟ ان سوالات کے واضح جواب کسی نئے کھیل کو محض ریاضیاتی دل چسپی سے آگے لے جاتے ہیں۔

\section{دو سے زیادہ کھلاڑی اور کوانٹم باہمی تعاون}
ہر کھیل دو کھلاڑیوں تک محدود نہیں ہوتا۔ معاشی منڈیاں، مواصلاتی شبکے، اجتماعی مذاکرات اور حیاتیاتی آبادیوں میں کئی عامل بیک وقت فیصلہ کرتے ہیں۔ تین یا زیادہ کھلاڑیوں کے لیے تدبیری فضا، ابتدائی مشترکہ کیفیت اور ہر کھلاڑی کی افادیت الگ الگ متعین کرنا ضروری ہے۔ بعض کوانٹم ماڈلوں میں کئی کھلاڑیوں کے مشترکہ کوانٹم وسائل ایسے توازن پیدا کرتے ہیں جن کی دو کھلاڑیوں والی مثال میں براہِ راست نظیر نہیں ملتی \cite{BenjaminHaydenMultiplayer}۔

تاہم ``تعاون'' کا لفظ یہاں احتیاط سے استعمال ہونا چاہیے: تجربہ گاہ میں ایک خاص مشترکہ کیفیت بانٹنا سماجی اعتماد، اخلاقی یک جہتی یا سیاسی اتفاق کے برابر نہیں۔ اسی طرح کسی ریاضیاتی ماڈل میں باہمی فائدہ ملنے سے یہ ثابت نہیں ہوتا کہ اس کے انسانی شرکا وہی فیصلہ کریں گے۔ ان دونوں سطحوں کو ملانا قابلِ فہم تمثیل تو ہوسکتا ہے، سائنسی دلیل نہیں۔

\section{پاکستان اور اردو خواں دنیا کے لیے علمی اہمیت}
اردو میں جدید طبیعیات بیان کرنے کا بنیادی فائدہ محض انگریزی اصطلاحات کا ترجمہ نہیں؛ اصل فائدہ یہ ہے کہ طالب علم اپنے سوالات، مثالیں اور استدلال اپنی علمی زبان میں منظم کرسکے۔ اس مقصد کے لیے چند طریقے قابلِ عمل ہیں۔ پہلی بات یہ کہ ہر نئے تصور کا اردو نام اور انگریزی اصطلاح ساتھ دی جائے، جیسا کہ اس مضمون میں \textbf{افادیت} \textenglish{(payoff)}، \textbf{نیش توازن} \textenglish{(Nash equilibrium)} اور \textbf{ارتقائی استحکام} \textenglish{(evolutionary stability)} کے ساتھ کیا گیا ہے۔ دوسری بات یہ کہ ایک ہی اصطلاح مختلف ابواب میں مختلف معنوں میں استعمال نہ ہو۔ تیسری بات یہ کہ اعداد، تجرباتی دعوے اور ادبی تمثیلیں الگ الگ پہچانی جاسکیں۔

اردو میں اس موضوع کی تدریس بین الشعبہ جاتی گفتگو کے لیے بھی موزوں ہے۔ حیاتیات کے طالب علم باز اور فاختہ کی مثال سے آبادی کا ارتقا سمجھ سکتے ہیں، معاشیات کے طالب علم قیدیوں کے مخمصے سے مراعات کے تضاد کو دیکھ سکتے ہیں، اور طبیعیات کے طالب علم مشترکہ کوانٹم کیفیت، عمل اور پیمائش کے ریاضیاتی تعلق پر غور کرسکتے ہیں۔ یہ سب \textbf{تعلیمی امکانات} ہیں؛ انہیں حقیقی سماجی مسائل کے خودکار کوانٹم حل کے دعوے میں تبدیل نہیں کرنا چاہیے۔

\section{اردو--انگریزی اصطلاحات: ایک مختصر فرہنگ}
\begin{center}
\renewcommand{\arraystretch}{1.22}
\begin{tabular}{p{0.45\linewidth}|p{0.42\linewidth}}
\textbf{اردو اصطلاح} & \textbf{انگریزی اصطلاح} \\ \hline
نظریۂ تدبیر / نظریۂ کھیل & \textenglish{Game theory} \\
کھلاڑی یا عامل & \textenglish{Player / agent} \\
تدبیر / حکمتِ عملی & \textenglish{Strategy} \\
افادیت / حاصل & \textenglish{Payoff / utility} \\
بہترین ردِعمل & \textenglish{Best response} \\
نیش توازن & \textenglish{Nash equilibrium} \\
خالص اور مخلوط تدبیر & \textenglish{Pure and mixed strategy} \\
ارتقائی طور پر مستحکم تدبیر & \textenglish{Evolutionarily stable strategy} \\
کیوبٹ & \textenglish{Qubit} \\
حالتوں کا خطی امتزاج & \textenglish{Superposition} \\
کوانٹم اشتراک / انٹینگلمنٹ & \textenglish{Entanglement} \\
یونیٹری عمل & \textenglish{Unitary operation} \\
کثافتی مصفوفہ & \textenglish{Density matrix} \\
پیمائش & \textenglish{Measurement} \\
ماحولیاتی بے ربطی & \textenglish{Decoherence} \\
ریپلیکیٹر حرکیات & \textenglish{Replicator dynamics} \\
\end{tabular}
\end{center}
یہ فرہنگ حتمی لغوی فیصلے کے طور پر نہیں بلکہ ایک قابلِ بحث تدریسی تجویز کے طور پر پیش کی گئی ہے۔ جہاں کسی لفظ کے کئی درست اردو متبادل ہوں، انگریزی اصطلاح درج رکھنے سے تحقیق و مطالعے میں سہولت رہتی ہے۔

\section{خود جانچ کے سوالات اور مختصر رہنمائی}
\begin{enumerate}[label=\arabic*.]
\item قیدیوں کے مخمصے کی مذکورہ جدول میں $(C,C)$ ہر کھلاڑی کو زیادہ افادیت دیتا ہے؛ پھر یہ نیش توازن کیوں نہیں؟ \textbf{رہنمائی:} ایک کھلاڑی کے یک طرفہ انحراف کا حاصل تقابل کریں۔
\item اگر $|\psi\rangle=\sqrt{3}/2|0\rangle+1/2|1\rangle$ ہو تو حسابی بنیاد میں صفر کی پیمائش کا امکان کیا ہے؟ \textbf{جواب:} $3/4$۔
\item باز--فاختہ کھیل میں $V=3$ اور $C=6$ ہو تو اندرونی توازن میں بازوں کا تناسب کیا ہوگا؟ \textbf{جواب:} $x^*=1/2$۔
\item صرف دو کھلاڑیوں کا ہم آہنگ نتیجہ دیکھ کر انٹینگلمنٹ کیوں ثابت نہیں ہوجاتی؟ \textbf{رہنمائی:} مشترکہ کلاسیکی تصادف کے ساتھ موازنہ کریں۔
\item کسی نئے کوانٹم توازن کا دعویٰ پڑھتے وقت تدبیری جماعت اور پیمائش کی صراحت کیوں ضروری ہے؟ \textbf{رہنمائی:} بہترین ردِعمل اس بات پر منحصر ہے کہ انحراف کرنے والے کھلاڑی کو کن اعمال کی اجازت ہے۔
\end{enumerate}
یہ سوالات ریاضیاتی عمل کی مشق کے ساتھ ساتھ ایک اہم سائنسی عادت پیدا کرتے ہیں: ہر دل چسپ نتیجے کے پیچھے موجود مفروضوں کی فہرست سامنے رکھی جائے۔

\section{ریاضیاتی بنیاد: کھیل، افادیت اور توازن کی تین سطحیں}
اس بحث میں سب سے اہم احتیاط یہ ہے کہ ایک ہی لفظ ``استحکام'' کو تین مختلف مفاہیم میں استعمال نہ کیا جائے۔ ایک کھیل کی تعیین کے لیے کھلاڑیوں کی تعداد، ہر کھلاڑی کی قابلِ اجازت تدابیر کا مجموعہ، اور تدبیروں کے ہر مجموعے کے لیے افادیت کا قاعدہ ضروری ہے۔ دو کھلاڑیوں کے کھیل کو مختصراً
\begin{equation}\label{eq:normal-form-deep}
G=(S_A,S_B;P_A,P_B)
\end{equation}
لکھتے ہیں۔ یہاں $S_A,S_B$ صرف ناموں کی فہرست نہیں بلکہ وہ \textbf{تمام} اعمال ہیں جنہیں قواعد کے تحت اختیار کیا جاسکتا ہے۔ اگر انہی قواعد کو بدل کر حکمت عملیوں کا مجموعہ وسیع کردیا جائے تو توازن کے متعلق فیصلہ بھی بدل سکتا ہے، خواہ افادیت کے عدد جوں کے توں رہیں۔

متشاکل کھیل میں $P(x,y)$ سے مراد اس کھلاڑی کی افادیت ہے جو $x$ اختیار کرے جبکہ مقابل $y$ اختیار کرتا ہو۔ تدبیر $s$ ایک متشاکل نیش توازن تب ہے جب
\begin{equation}\label{eq:nash-deep}
P(s,s)\ge P(t,s)\qquad \text{ہر }t\in S.
\end{equation}
اگر $t\ne s$ کے لیے ہر مرتبہ سخت عدم مساوات ہو تو یہ \textbf{سخت نیش توازن} ہے۔ اس تعریف میں ایک فرد کی یک طرفہ تبدیلی کا سوال ہے؛ اس سے ابھی یہ ثابت نہیں ہوتا کہ آبادی میں ایک نئی تدبیر پھیل نہیں سکے گی۔

ارتقائی طور پر مستحکم تدبیر (\textenglish{ESS}) کی مینا رڈ اسمتھ والی شرط نسبتاً زیادہ تقاضا کرتی ہے: ہر $t\ne s$ کے لیے یا تو
\begin{equation}\label{eq:ess-a-deep}
P(s,s)>P(t,s),
\end{equation}
یا، پہلی مساوات کی صورت میں،
\begin{equation}\label{eq:ess-b-deep}
P(s,s)=P(t,s)\quad\text{اور}\quad P(s,t)>P(t,t).
\end{equation}
اس کی وجہ مختصر حساب سے سمجھ میں آتی ہے۔ آبادی میں $s$ کی کثرت $1-\varepsilon$ اور متغیر $t$ کی کثرت $\varepsilon$ ہو تو
\begin{align}
F_s(\varepsilon)&=(1-\varepsilon)P(s,s)+\varepsilon P(s,t),\\
F_t(\varepsilon)&=(1-\varepsilon)P(t,s)+\varepsilon P(t,t).
\end{align}
دونوں میں فرق
\begin{equation}\label{eq:invasion-deep}
F_s-F_t=(1-\varepsilon)[P(s,s)-P(t,s)]
+\varepsilon[P(s,t)-P(t,t)]
\end{equation}
ہے۔ پہلی شرط کے تحت کم تعداد میں داخل ہونے والا متغیر شکست کھاتا ہے؛ پہلی برابری کی صورت میں دوسری شرط فیصلہ کرتی ہے۔ یہ نتیجہ دو خالص اقسام کی نادر دراندازی کے لیے ہے۔ غیر محدود یا مسلسل تدبیری مجموعے میں الگ الگ متغیروں کے لیے حاصل کردہ چھوٹی حدیں لازماً ایک مشترک مثبت حد نہیں بناتیں۔ اسی لیے \textbf{ہر متغیر کے لیے} حملے کی روک تھام اور \textbf{تمام متغیروں کے لیے یکساں} روک تھام میں امتیاز ضروری ہے \cite{MaynardSmith,Weibull,IqbalAbbott2026}۔

\subsection{ایک اہم اصلاح: ریپلیکیٹر کے ہر ساکن نقطے کو ESS نہیں کہا جاسکتا}
ریپلیکیٹر حرکیات میں ہر اس نقطے پر جہاں متعلقہ اقسام کی تناسبی بڑھوتری صفر ہے، $\dot x=0$ ہوسکتا ہے۔ تاہم ایک ساکن نقطہ غیر مستحکم بھی ہوسکتا ہے؛ محض ساکن ہونے سے ارتقائی استحکام ثابت نہیں ہوتا۔ مناسب مفروضوں کے تحت ESS اور حرکیاتی استحکام کے درمیان گہرا رشتہ ہوتا ہے، مگر عمومی مساوات نہیں۔ اسی سبب گزشتہ متن میں ایسی عبارت جو ESS کو محض ریپلیکیٹر کے ساکن نقاط کے برابر قرار دیتی تھی، ریاضیاتی لحاظ سے ضرورت سے زیادہ وسیع تھی؛ درست بیان یہ ہے کہ \textbf{ESS کا حرکیاتی نتیجہ ماڈل اور استحکام کی منتخب تعریف پر منحصر ہے} \cite{Hofbauer,Cressman}۔

\section{ایک مکمل کلاسیکی مثال: توازن تو ہے، ارتقائی استحکام نہیں}
دو تدبیری اشتراکی کھیل (\textenglish{coordination game}) کا میٹرکس فرض کیجیے:
\begin{equation}\label{eq:coord-matrix}
A=\begin{pmatrix}a&b\\b&a\end{pmatrix},\qquad a>b.
\end{equation}
ایک ہی تدبیر اختیار کرنے والے دونوں کھلاڑیوں کو زیادہ افادیت ملتی ہے۔ اگر پہلے کھلاڑی کے پہلی تدبیر اختیار کرنے کا امکان $p$ اور دوسرے کا $q$ ہو تو
\begin{equation}\label{eq:coord-payoff}
P(p,q)=a[pq+(1-p)(1-q)]+b[p(1-q)+(1-p)q].
\end{equation}
$p^*=1/2$ پر مقابل کی کسی بھی مخلوط تدبیر $q$ کے خلاف افادیت $(a+b)/2$ ہے۔ یوں $p^*$ نیش توازن ہے کیونکہ یک طرفہ تبدیلی سے زیادہ افادیت نہیں ملتی۔ البتہ $q\ne1/2$ کے لیے
\begin{align}
P(p^*,q)&=\frac{a+b}{2},\\
P(q,q)&=\frac{a+b}{2}+2(a-b)\left(q-\frac12\right)^2,
\end{align}
لہٰذا
\begin{equation}\label{eq:coord-notess}
P(p^*,q)-P(q,q)=-2(a-b)\left(q-\frac12\right)^2<0.
\end{equation}
پہلی ESS شرط پر برابری ہے لیکن دوسری شرط الٹی سمت میں جاتی ہے؛ اس لیے اندرونی نیش توازن ESS نہیں۔ اس کے برعکس دونوں خالص اشتراکی تدبیریں سخت نیش توازن ہیں۔

آبادی میں پہلی تدبیر کا تناسب $x$ مانیں۔ دونوں خالص تدبیروں کی افادیت کا فرق $(a-b)(2x-1)$ ہے، اس لیے
\begin{equation}\label{eq:coord-replicator}
\dot x=(a-b)x(1-x)(2x-1).
\end{equation}
$x=1/2$ کے بائیں جانب $\dot x<0$، اور دائیں جانب $\dot x>0$؛ لہٰذا یہ نقطہ \textbf{غیر مستحکم} ہے، اگرچہ نیش توازن ہے۔ یہ سادہ مثال واضح کرتی ہے کہ محض توازن کا مل جانا کافی نہیں۔ تازہ کوانٹم تحقیق میں اسی طرز کے اشتراکی کھیل کو بنیاد بنا کر محدود اور وسیع تدبیری دائروں کا فرق کھولا گیا ہے \cite{IqbalAbbott2026}۔

\section{کوانٹم کھیل کی ٹھوس تعریف: عمل، پیمائش اور افادیت}
کوانٹم کھیل میں کلاسیکی افادیت کے جدول کو بدلنا لازمی نہیں؛ بدلنے والی چیز ممکنہ نتائج کی تقسیم اور ان پر کھلاڑیوں کا تصرف ہوسکتی ہے۔ آیزرٹ، ولکنز اور لیونسٹائن (\textenglish{EWL}) کے معروف طریقے میں دو کیوبٹ ابتدائی حالت $|00\rangle$ سے ایک مقرر شدہ دروازہ $J$ عمل کرتا ہے۔ اس کے بعد کھلاڑی اپنے مقامی اعمال $U_A$ اور $U_B$ اختیار کرتے ہیں، اور آخر میں $J^\dagger$ کے بعد پیمائش ہوتی ہے:
\begin{equation}\label{eq:ewl-deep}
|\psi_f\rangle=J^\dagger(U_A\otimes U_B)J|00\rangle.
\end{equation}
مثلاً ایک استعمال شدہ کنونشن میں
\begin{equation}\label{eq:J-deep}
J(\gamma)=\exp\!\left(\frac{i\gamma}{2}\,\sigma_x\otimes\sigma_x\right),
\quad 0\le\gamma\le\frac{\pi}{2}.
\end{equation}
$\gamma$ ابتدا میں بننے والے اشتراک کی شدت کو کنٹرول کرتا ہے۔ دیگر مصنفین مختلف لیکن مناسب طور پر مساوی کنونشن استعمال کرتے ہیں؛ ان کا تقابل کرتے وقت علامات اور قواعد کی مطابقت ضروری ہے۔ مخصوص مطابقت پذیر کلاسیکی اعمال منتخب ہوں تو اصل کلاسیکی کھیل اس ماڈل کے اندر واپس مل سکتا ہے، مگر \textbf{ہر} من مانے یونٹری انتخاب کو کلاسیکی تدبیر کہنا درست نہیں۔

افادیت صرف حالت کے لکھے ہوئے امپلیٹیوڈ کا نام نہیں۔ ہر نتیجے $ij$ کے لیے $p_{ij}=|\langle ij|\psi_f\rangle|^2$ معلوم کرکے
\begin{equation}\label{eq:observable-deep}
P_A=\operatorname{Tr}(R_A\rho_f),\qquad
R_A=\sum_{ij}a_{ij}|ij\rangle\langle ij|,
\quad \rho_f=|\psi_f\rangle\langle\psi_f|
\end{equation}
حاصل ہوگا؛ یہی اصول دوسرے کھلاڑی کے لیے بھی ہے۔ اس صورت میں نتیجہ یکسر واضح ہے: کسی تدبیر کو ``فاتح'' یا ``ارتقائی طور پر مستحکم'' کہنے سے پہلے مکمل افادیتی عامل $R_A$، ابتدائی حالت، انجام دیے گئے اعمال، اور پیمائش بتانا ضروری ہے۔

\subsection{اختیاری تدبیروں کی تحدید کیوں بنیادی اہمیت رکھتی ہے؟}
$SU(2)$ ایسے دو بُعدی یونٹری میٹرکس کا مجموعہ ہے جن کا ڈیٹرمیننٹ ایک ہے۔ EWL کے بہت سے ابتدائی تجزیوں میں اس کی ایک دو پارامیٹر والی ذیلی فیملی منتخب کی گئی۔ اس میں کسی تدبیر کا دوسرے انتخاب سے محفوظ رہنا اس بات کی ضمانت نہیں کہ وہ \textbf{پورے} $SU(2)$ میں بھی محفوظ ہو۔ اس اصول کو بینجمن اور ہیڈن کے تبصرے نے نمایاں کیا، اور بعد کی بحث میں یہ مرکزی احتیاط بنا رہا \cite{BenjaminHaydenComment,EisertReply,KhanReview2018}۔

اس کا عمومی ریاضیاتی اظہار بہت سادہ ہے: اگر $S_0\subset S_1$ اور $s\in S_0$ ہو، تو
\begin{equation}
P(s,s)\ge P(t,s)\quad(t\in S_0)
\end{equation}
سے یہ نتیجہ \textbf{نہیں} نکلتا کہ یہی عدم مساوات $t\in S_1$ کے لیے بھی درست ہوگی۔ کلاسیکی کھیل میں نئی تدبیر شامل کرنے سے بھی ایسا ہوسکتا ہے؛ اس لیے ہر تبدیلی کو کوانٹم اسرار کہنا مناسب نہیں۔

\section{مساوی افادیت: دو مختلف یونٹری اعمال کب ایک ہی تدبیر ہیں؟}
اگر دو یونٹری اعمال $U,V$ بظاہر مختلف ہوں لیکن ہر ممکن حریف $W$ کے مقابل ان سے حاصل ہونے والی افادیت ایک جیسی ہو، تو ارتقائی ماڈل میں سوال پیدا ہوتا ہے: کیا ان اعمال کو دو جدا اقسام سمجھنا چاہیے؟ ایک مضبوط تعریف میں $U\sim V$ تب لکھیں گے جب \textbf{ہر} حریف کے مقابل، دونوں کھلاڑیوں کے تمام متعلقہ کرداروں میں، ان کے افادیتی نتائج برابر ہوں۔ صرف کسی ایک مقررہ حریف کے سامنے اتفاقاً برابر افادیت مل جانا کافی نہیں۔

اس سوال کی خاص اہمیت اس اشتراکی کھیل میں ہے جہاں
\begin{equation}\label{eq:agree-deep}
a_{00}=a_{11}=a,\qquad a_{01}=a_{10}=b.
\end{equation}
اس وقت
\begin{equation}\label{eq:agree-payoff-deep}
P=b+(a-b)(p_{00}+p_{11}).
\end{equation}
یعنی پوری افادیت صرف اتفاق والے دو نتائج کے مجموعی امکان پر منحصر ہے۔ کسی دوسرے عمومی افادیتی جدول میں یہ سہولت موجود نہیں ہوگی۔ اسی لیے یہاں حاصل ہونے والی مساوی افادیت کی درجہ بندی کو ہر کوانٹم کھیل پر چسپاں نہیں کیا جاسکتا۔

\subsection{بلوخ کرہ اور $SU(2)$ کی زائد نمائندگی}
پاؤلی میٹرکس $\sigma_z$ لے کر ہر $U\in SU(2)$ کے ساتھ درج ذیل تعلق منسوب کیا جاسکتا ہے:
\begin{equation}\label{eq:bloch-deep}
U^\dagger\sigma_z U=n_x(U)\sigma_x+n_y(U)\sigma_y+n_z(U)\sigma_z,
\qquad |\mathbf n(U)|=1.
\end{equation}
یوں $\mathbf n(U)$ سہ بُعدی اکائی کرہ (بلوخ کرہ) کا ایک نقطہ ہے۔ اس مخصوص اشتراکی کھیل میں افادیت یونٹری میٹرکس کے تمام اجزا کی بجائے اس کے متعلقہ بلوخ سمتیے سے طے ہوتی ہے۔ $V_\phi=\operatorname{diag}(e^{i\phi},e^{-i\phi})$ کے لیے $V_\phi^\dagger\sigma_zV_\phi=\sigma_z$، اس لیے
\begin{equation}\label{eq:coset-deep}
\mathbf n(V_\phi U)=\mathbf n(U).
\end{equation}
یہ محض حسابی اتفاق نہیں بلکہ ایک پورے $U(1)$ ذیلی گروہ کی غیر قابلِ تمیزیت ہے۔ مثبت انٹینگلمنٹ اور مذکورہ مخصوص افادیتی ساخت میں اسی سے quotient
\begin{equation}\label{eq:quotient-deep}
U(1)\backslash SU(2)\cong S^2
\end{equation}
حاصل ہوتا ہے۔ اس نتیجے کا مفصل اثبات حالیہ تحقیقی مقالے میں موجود ہے؛ یہاں اس کی ریاضیاتی منطق اور تعبیر پیش کی جارہی ہے، اسے نیا مستقل قضیہ قرار نہیں دیا جا رہا \cite{IqbalAbbott2026}۔

دو تعبیرات سامنے آتی ہیں۔ \textbf{غیر مختزل ماڈل} میں ہر یونٹری ایک جدا ارتقائی قسم ہے؛ افادیتی طور پر ہم مثل مختلف یونٹری اعمال کی موجودگی سخت ESS کے دعوے میں رکاوٹ بن سکتی ہے۔ \textbf{مختزل ماڈل} میں ہم مثل اعمال ایک ہی نقطہ شمار ہوتے ہیں، اور متعلقہ مثال میں مثبت انٹینگلمنٹ کے لیے سخت توازن و ESS بحال ہوسکتے ہیں۔ دونوں نتائج میں تضاد نہیں؛ دونوں کا تدبیری مجموعہ مختلف ہے۔ اگر طبیعی ہارڈویئر میں مساوی افادیت والے اعمال کی الگ لاگت، شور یا توانائی مطلوب ہو تو انہیں ایک ہی قسم ماننے کا جواز الگ سے جانچنا ہوگا۔

\subsection{کیا ہر دو مساوی افادیت والے اعمال ایک دوسرے کے خلاف غیر جانب دار ہیں؟}
اگر دو تدبیروں سے ہر حریف کے سامنے برابر افادیت ملتی ہے، تو خالص کھیل کی افادیتی تعریف کے لحاظ سے انتخاب ان کے مابین فرق نہیں کرسکتا۔ تاہم یہ \textbf{غیر جانب داری}، \textbf{ارتقائی استحکام} سے مختلف ہے۔ افادیتی مساوی جماعت (\textenglish{payoff-equivalence class}) کو بحیثیتِ مجموعہ حملہ آوروں سے محفوظ رکھنا ممکن ہے، جبکہ اس جماعت کے ہر نمائندہ یونٹری کو علیحدہ سخت ESS کہنا غلط ہو۔ اس فرق سے \textenglish{evolutionarily stable sets} اور غیر جانب دار سمتی بہاؤ کی طرف مطالعے کا راستہ کھلتا ہے \cite{IqbalAbbott2026}۔

\section{حساسیت اور مضبوطی: معمولی تبدیلی کب نتیجہ بدل دیتی ہے؟}
ایک سائنسی نتیجے کی اہمیت صرف اس کے درست ہونے میں نہیں بلکہ اس میں بھی ہے کہ مفروضوں کی چھوٹی تبدیلی پر وہ کہاں تک برقرار رہتا ہے۔ اگر کھیل کی افادیت کا انحصار صرف $p_{00}+p_{11}$ پر ہے تو $a_{00}=a_{11}$ اور $a_{01}=a_{10}$ کی مساوات محوری ہے۔ فرض کیجیے کسی حقیقی تجربے میں ان میں سے ایک ادائیگی کو $\delta$ سے بدل دیا گیا۔ افادیت کے نئے عامل میں ایک اضافی اصطلاح
\begin{equation}\label{eq:perturb-deep}
R_A'=R_A+\delta|00\rangle\langle00|
\end{equation}
آجائے گی۔ پھر
\begin{equation}
P_A'-P_A=\delta p_{00},\qquad |P_A'-P_A|\le |\delta|.
\end{equation}
یہ حد بتاتی ہے کہ \textbf{افادیت میں عددی تبدیلی} چھوٹی ہے، لیکن یہ ضمانت نہیں کہ زیادہ سے زیادہ افادیت دینے والی تدبیر یا ESS کی \textbf{شناخت} بھی غیر متغیر رہے گی۔ اگر سابقہ توازن میں بعض انحرافات کے خلاف سخت فاصلہ صفر یا نہایت چھوٹا ہو تو معمولی $\delta$ بھی توازن کی نوعیت بدل سکتا ہے۔ مذکورہ خاص کھیل میں یہ نازک حساسیت نئی تحقیق کی نمایاں حد ہے \cite{IqbalAbbott2026}۔

\subsection{شور کے ساتھ عمومی کثافت میٹرکس}
حقیقی تجربوں میں خالص حالت کے بجائے کثافت میٹرکس $\rho$ کام آتا ہے۔ اگر $\mathcal E$ مکمل مثبت، ٹریس محفوظ (\textenglish{CPTP}) نقشہ ہو تو
\begin{equation}\label{eq:noise-deep}
\rho_f=J^\dagger\,\mathcal E\!\left[(U_A\otimes U_B)J\rho_0J^\dagger(U_A^\dagger\otimes U_B^\dagger)\right]J.
\end{equation}
$\mathcal E$ شور، نقص یا کسی متعین تجرباتی عمل کی نمائندگی کرسکتا ہے؛ اس کی جگہ اور ترتیب واضح کرنا ضروری ہے۔ پھر $P_A=\operatorname{Tr}(R_A\rho_f)$ ہوگا۔ دو اختتامی حالتوں $\rho_f,\widetilde\rho_f$ کے لیے
\begin{equation}\label{eq:trace-bound-deep}
|\operatorname{Tr}[R_A(\rho_f-\widetilde\rho_f)]|
\le \|R_A\|_\infty\,\|\rho_f-\widetilde\rho_f\|_1.
\end{equation}
یہ عمومی حد تجرباتی غلطی سے افادیت کی غلطی کو باندھتی ہے۔ اس میں $\|\cdot\|_1$ ٹریس نورم اور $\|\cdot\|_\infty$ آپریٹر نورم ہے۔ پھر بھی ESS کی مضبوطی جانچنے کے لیے تمام قابلِ اجازت متغیر تدبیروں کے مقابل ادائیگی کے فرق کا تجزیہ کرنا ہوگا، محض ایک یا دو عددی تجربات کافی نہیں۔

\section{آبادی کی حرکیات اور کوانٹم ارتقا کی تعبیر}
کوانٹم کھیل کے افادیتی اعداد کو آبادی کی حرکیات میں شامل کرنا دو جدا نظریوں کا اتصال ہے۔ فرض کریں $n$ وراثت پذیر اقسام ہیں اور ان کے تناسب $x_i$ ہیں۔ عمومی ریپلیکیٹر مساوات
\begin{equation}\label{eq:replicator-many-deep}
\dot x_i=x_i[f_i(\mathbf x)-\bar f(\mathbf x)],\qquad
\bar f(\mathbf x)=\sum_j x_j f_j(\mathbf x),\quad\sum_i x_i=1
\end{equation}
ہے۔ اگر دو اقسام کی جوڑیوں کے مقابلوں کی افادیت کوانٹم پروٹوکول سے پیدا ہو تو ان سے $f_i$ حساب کیا جاسکتا ہے۔ \textbf{لیکن} اس سے یہ ثابت نہیں ہوتا کہ حیاتیاتی آبادی واقعی مربوط کوانٹم حالتوں میں رہتی ہے۔ یہ ایک مفروضاتی کھیل و ارتقا کا ماڈل بھی ہوسکتا ہے جس میں کوانٹم آلہ فیصلہ سازی کے نتائج پیدا کرتا ہے۔

اس ربط کی جانچ کے لیے پوچھنا چاہیے: ہر مقابلے سے پہلے نئی کوانٹم حالت کون تیار کرتا ہے؟ کیا جوڑیوں کے درمیان مشترکہ رابطہ ہے؟ کیا انواع کی افزائش کے وقت تدبیری معلومات نقل ہوتی ہے یا کیوبٹ کی حالت؟ کیا ماحول ڈی کوہیرنس پیدا کرتا ہے؟ اور کیا مختلف یونٹری نمائندوں کی افزائش یکساں ہے؟ ان سوالوں کے جواب کے بغیر کسی حیاتیاتی عمل میں ``کوانٹم ڈارونزم'' یا ``کوانٹم ESS'' کا حقیقی وجود ثابت نہیں ہوتا۔

\subsection{محدود آبادی میں اتفاقی انحراف}
ریپلیکیٹر مساوات لامحدود یا بہت بڑی آبادی کی مثالی حد ہے۔ چھوٹی آبادی میں اتفاقی تبدیلی (\textenglish{genetic drift}) نادر متغیر کے ختم ہونے یا غالب آنے پر اثر انداز ہوسکتی ہے۔ اس لیے ``حملہ آور کامیاب نہیں ہوسکتا'' کی نظریاتی عبارت کو \textbf{یقینی تجرباتی ناممکن} نہیں سمجھنا چاہیے۔ محدود آبادی میں موزوں سوال کسی متغیر کے تثبیت پانے کے امکان (\textenglish{fixation probability}) سے بھی متعلق ہوگا۔ یہ کلاسیکی ارتقائی احتیاط کوانٹم سے حاصل افادیت رکھنے والے ماڈل پر بھی منطبق ہوتی ہے۔

\section{کوانٹم برتری اور بیل عدم مساوات: متعلق مگر الگ سوال}
کوانٹم کھیل میں بہتر مجموعی افادیت کا حصول اور بیل عدم مساوات کی خلاف ورزی ایک ہی بات نہیں۔ پہلی چیز کھیل کی افادیتی ساخت، توازن اور قابلِ اجازت تدبیروں پر منحصر ہے؛ دوسری مخصوص غیر مقامی تعلقات کے تجرباتی تقابل سے متعلق ہے۔ مثال کے طور پر معیاری \textenglish{CHSH} منظرنامے میں دو دور افتادہ فریق دو دو پیمائشی انتخاب رکھتے ہیں۔ کلاسیکی مقامی مخفی متغیر ماڈل میں
\begin{equation}\label{eq:chsh-deep}
|E_{00}+E_{01}+E_{10}-E_{11}|\le2,
\end{equation}
جبکہ مناسب کوانٹم پیمائشوں سے $2\sqrt2$ تک قدر مل سکتی ہے۔ یہ کوانٹم تعلقات کی خاص نوعیت کو ظاہر کرتا ہے، لیکن عام EWL کھیل میں کسی کھلاڑی کی بہتر افادیت سے خودبخود \textenglish{CHSH} کی خلاف ورزی ثابت نہیں ہوتی۔ منصفانہ دعوے کے لیے مشترکہ کلاسیکی بے ترتیبی، مواصلاتی وسائل، کھلاڑیوں کی معلومات اور دستیاب پیمائشوں کو یکساں بنیاد پر رکھنا ہوگا \cite{NielsenChuang,KhanReview2018}۔

\section{ایک قابلِ تکرار تحقیقی طریقۂ کار اور آئندہ کا عملی ایجنڈا}
تحقیق کے لیے ایک مفید معیار یہ ہے کہ محض نتیجہ نہیں بلکہ اسے دوبارہ حاصل کرنے کا مکمل طریقہ شائع کیا جائے۔ ہر نئے کوانٹم کھیل کے لیے درج ذیل نکات بطور جانچ فہرست پیش کیے جاسکتے ہیں:
\begin{enumerate}[label=\arabic*.]
\item کھیل کے کلاسیکی افادیتی میٹرکس، متشاکل یا غیر متشاکل نوعیت اور کھلاڑیوں کے معلوماتی مفروضے متعین ہوں۔
\item $\rho_0$، انٹینگلنگ عمل، قابلِ اجازت مقامی یا عمومی عملیات، پیمائش اور افادیتی عوامل واضح لکھے جائیں۔
\item ثابت کیا جائے کہ منتخب کلاسیکی اعمال سے کلاسیکی کھیل واپس ملتا ہے؛ یہ مفروضہ محض بیان نہ کیا جائے۔
\item نیش کی شرط کو \textbf{مکمل} تدبیری مجموعے پر جانچا جائے، نہ کہ صرف چند منتخب مثالوں پر۔
\item ESS کی پہلی اور دوسری شرائط، مساوی افادیت والی اقسام اور مختزل بمقابلہ غیر مختزل دائرے الگ الگ رپورٹ ہوں۔
\item ریپلیکیٹر حرکیات کے ساکن نقاط کے ساتھ ان کی مقامی و عمومی استحکامی جانچ بھی شامل ہو۔
\item نتائج کی حساسیت $\gamma$، شور، افادیت کی غیر یقینی کیفیت اور چھوٹی تبدیلیوں کے مقابل آزمائی جائے۔
\item اگر ``کوانٹم فائدہ'' کا دعویٰ کیا جائے تو مساوی وسائل والا ایک واضح کلاسیکی تقابلی طریقہ دیا جائے۔
\end{enumerate}

یہ فہرست کسی نئی تھیوری کا ثبوت نہیں بلکہ تحقیق کی صداقت اور قابلِ تکرار ہونے کے لیے مجوزہ طریقہ ہے۔ عملی تحقیق کے لیے تین نسبتاً زیادہ نتیجہ خیز راستے سامنے آتے ہیں: (الف) عام افادیتی عوامل کے لیے مساوی افادیت کی جیومیٹری کو سمجھنا، (ب) شور اور تدبیری خطاؤں کے تحت سخت توازن کی مضبوطی کو مقداری حدوں میں بیان کرنا، اور (ج) محدود آبادیوں میں کوانٹم پروٹوکول کی افادیت سے پیدا ہونے والی اتفاقی ارتقائی حرکیات کو جانچنا۔ ان تینوں میں ریاضیاتی ثبوت، عددی محاکات اور تجرباتی معنی الگ الگ بیان کیے جانے چاہئیں۔

\section{تنقیدی نتیجہ: نئے امکانات کے ساتھ بنیادی حدود کی شناخت}
اس مطالعے سے یہ سبق سامنے آتا ہے کہ کوانٹم نظریۂ تدبیر کو کلاسیکی نظریے کی محض مشکل تر زبان قرار دینا بھی ناکافی ہے اور اسے ہر مخمصے کا خودکار حل سمجھنا بھی درست نہیں۔ ایک کوانٹم پروٹوکول تدبیری امکانات، نتیجے کی تقسیم اور معلوماتی وسائل میں نئی صورتیں پیدا کرسکتا ہے؛ اس سے ملنے والا مخصوص توازن البتہ ہمیشہ پروٹوکول کی تعریف، قابلِ اجازت اعمال، افادیتی مشاہدہ اور متغیر تدبیروں کی شناخت کے تابع ہوگا۔ خصوصاً اشتراکی کھیلوں میں بلوخ کرہ اور مساوی افادیت کی جماعتوں کے ذریعے سامنے آنے والی ساخت ہمیں بتاتی ہے کہ \textbf{ایک الگ یونٹری کا استحکام} اور \textbf{افادیتی جماعت کا استحکام} دو جدا سائنسی دعوے ہیں۔

اردو قاری کے لیے اس علم کی اہمیت محض اصطلاحات کے ترجمے تک محدود نہیں: یہ جدید طبیعیات، ریاضیاتی حیاتیات، فیصلہ سازی اور فلسفۂ سائنس کے مابین ایسے سوالات سے واقف کراتا ہے جن کے جواب واضح تعریفوں، احتیاط سے کیے گئے حساب اور قابلِ تکرار تجربات کے بغیر نہیں مل سکتے۔ اسی تحقیقی دیانت سے نظریۂ تدبیر کے کوانٹم توسیعی پروگرام کی حقیقی قدر متعین ہوگی۔

\section{خلاصہ اور آئندہ تحقیق کی سمت}
کلاسیکی نظریۂ تدبیر باہمی فیصلوں کی منطق فراہم کرتا ہے؛ کوانٹم توسیع اس منطق میں نئی کیفیات، اعمال اور پیمائشوں کے ذریعے امکان کے نئے ڈھانچے شامل کرتی ہے۔ ان تبدیلیوں سے بعض متعین کھیلوں میں توازن اور استحکام بدل سکتے ہیں، مگر اس تبدیلی کی مقدار اور تعبیر قواعد و حدود کی پابند ہوتی ہے۔ آئندہ تحقیق کے لیے خاص طور پر تین باتیں اہم ہیں: مختلف تدبیری جماعتوں کے تحت نتائج کی مضبوطی، افادی طور پر مماثل تدابیر کی درست شناخت، اور تجرباتی شور کے تحت ارتقائی نتائج کی پائیداری۔ یہی امتیازات اردو میں ایک واضح اور محتاط سائنسی گفتگو کی بنیاد بن سکتے ہیں۔

\section{کوانٹم نظريۂ تدبير کے کچھ جوابات کے متلاشی سوالات}
تدبير کے کوانٹم نظریہ میں کئی سوالات ايسے موجود ہيں جو اب تک جوابات کے متلاشی ہيں۔ مثال کے طور پر
:
\\
1 ۔
کيا واقعی اور کس حد تک کوانٹم تدبيروں کو کلاسيکی تدبيروں کی قابل يقين توسيع سمجھا جا سکتا ہے؟
\\
2  ۔
کونسے حالات میں اور کن کھيلوں ميں کوانٹم تدبيريں ان کھيلوں کے کلاسیکی تدبيروں والے حل کو بھی اپنے اندر سمو ليتی ہيں ؟
\\
3 ۔
کیا ہم کلاسیکی کھیلوں  کو کوانٹم تدبيروں سے کھيلنے کے نئے طريقے معلوم کر سکتے ہيں يا معلوم کوانٹم تدبيروں ميں بہتری لا سکتے ہيں؟
\\
4 ۔
غير متحرّک اور متحرّک اقسام کے وہ کھيل
(\textenglish{static and dynamic games})
جن ميں کلاسيکی تدبيروں کی تعداد محدود ، گنی چنی، يا لا محدود ہوسکتی ہے، کو کوانٹم تدبيروں سے کيسے کھيلا جا سکتا ہے؟   
\\
5 ۔
کيا کوانٹم تدبيروں سے کھيلے گئے کھيل، امکانات کے کوانٹم نظريے
(\textenglish{theory of quantum probability})
کو بہتر طور پر سمجھنے ميں کوئی مدد کر سکتے ہيں؟ اگر ہاں تو کيسے؟
\\
6 ۔
کھيلوں کو کوانٹم تدبيروں سے کھيلنے کے طريقوں کے کوانٹم انفارميشن اور کمپيوٹيشن کے ميدان ميں کيا اطلاقات ہو سکتے ہيں؟ 

\section*{حوالہ جات}
\begingroup
\renewcommand{\theenumiv}{\number\value{enumiv}}
\makeatletter
\renewcommand{\@biblabel}[1]{[\number\c@enumiv]}
\makeatother
\begin{english}

\end{english}
\endgroup                                                                             
\section{مصنفّين کا تعارف}
\textbf{ادارتی نوٹ:} ذیل کا سوانحی خاکہ اصل مضمون کی تاریخی صورت کو بڑی حد تک برقرار رکھتا ہے۔ ادارہ جاتی مناصب، تعیناتیوں اور جرائد کی رکنیت سے متعلق زمانی الفاظ اصل بیان کے زمانے کے مطابق پڑھے جائیں؛ انہیں نومبر ۲۰۲۶ کی مصدقہ تازہ ترین فہرست نہ سمجھا جائے۔\par
\textbf{اظہر اقبال}
نے
1991
میں گورنمنٹ کالج لاہور پاکستان (جو اب گورنمنٹ کالج یونیورسٹی ہے) سے طبیعیات اور رياضی ميں بی ایس سی کی سند حاصل کی اور 1995 میں گورنمنٹ آف پاکستان کے سکالرشپ پریونیورسٹی آف شیفیلڈ  برطانیہ
(\textenglish{University of Sheffield, UK})
سے طبیعیات میں بی ایس سی (آنرز) کی ڈگری حاصل کی۔ وطن واپسی پر اس نے پاکستان انسٹی ٹیوٹ آف لیزرز اینڈ آپٹکس اسلام آباد پاکستان
(\textenglish{Pakistan Institute of Lasers and Optics, Islamabad, Pakistan})
میں فوٹوونکس
(\textenglish{photonics})
کے ميدان میں سال 2002  تک کام کیا۔ سال 2006  میں یونیورسٹی آف ھل
مشرقی یارک شائر برطانیہ
(\textenglish{University of Hull, East Yorkshire, UK})
سے اطلاقی ریاضيات
(\textenglish{Applied Mathematics})
میں پی ایچ ڈی کی ڈگری حاصل کرنے کے بعد اس نے نیشنل یونیورسٹی آف سائنسيز اور ٹیکنالوجی (\textenglish{NUST})
اسلام آباد پاکستان ميں تدريس کی۔  2006-2007  کے دوران جاپان سوسائٹی آف پروموشن آف سائينس
(\textenglish{JSPS})
کی طرف سے ايک پوسٹ ڈاکٹرل فيلوشپ
(\textenglish{postdoctoral fellowship})
 حاصل کرنے کے بعد اس نے کوچی يونيورسٹی آف ٹیکنالوجی
(\textenglish{Kochi University of Technology})
جنوبی جاپان ميں تحقيق کا کام کيا ۔ 2007-2011 کے دوران اس نے آسٹریلوی ریسرچ کونسل
(\textenglish{Australian Research Council})
یا
\textenglish{ARC}
 کی طرف سے دی گئی ايک پوسٹ  ڈاکٹرل فيلوشپ پر یونیورسٹی آف ایڈيلیلڈ
 (\textenglish{University of Adelaide})
 جنوبی آسٹريليا ميں تحقيق کا کام کيا ۔ 2012 کے سال  کے دوران اظہر نے یونیورسٹی آف ایڈيلیلڈ  ميں
 \textenglish{ARC}
کے فنڈز پر بطور سینئر ریسرچ ایسوسی ايٹ
 (\textenglish{Senior Research Associate})
 کام کيا۔ 2013 کے آغاز سے اس نے سعودی عرب کی شاہ فہد یونیورسٹی آف پٹرولیم اور منرلز دھران
  (\textenglish{King Fahd University of Petroleum and Minerals, Dhahran, Saudi Arabia })
کے رياضيات اور شماريات کے ڈيپارٹمنٹ
 (\textenglish{Department of Mathematics \& Statistics})
ميں تدريس کی۔ 2014 کے وسط سے وہ شاہ فہد یونیورسٹی آف پٹرولیم اور منرلز سے  یونیورسٹی آف ایڈيلیلڈ  واپس آگئے جہاں وہ تحقيق اور تدريس سے وابستہ ہيں۔ انکی نظرياتی طبیعيات اور اطلاقی رياضی ميں تحقیق کا قابل ذکرحصہ کوانٹم گیم تھیوری یا تدبير کے کوانٹم نظریے پر ہے اور انہيں بين الکلياتی تحقيق
 (\textenglish{interdisciplinary research})
سے بھی دلچسپی ہے۔

    \textbf{ڈيرک ايبٹ}
    نےبی ايس سی آنرز کی ڈگری يونيورسٹی
    آف لوفبوروہ برطانيہ
 (\textenglish{University of Loughborough, UK})
سے اور پی ايچ ڈی کی ڈگری يونيورسٹی آف ايڈيليڈ آسٹريليا سے اليکٹريکل اور اليکٹرانک انجينيئرنگ ميں حاصل کی۔
\textenglish{1978-1986}
کے دوران وہ جی ای سی ہرسٹ ريسرچ سنٹر لنڈن
(\textenglish{GEC Hirst Research Centre, London})
ميں ايک ريسرچ انجينيئر
(\textenglish{Research Engineer})
رہے۔
\textenglish{1986-1987}
کے دوران وہ آسٹک مائيکرو سسٹمز آسٹريليا
(\textenglish{Austek Microsystems, Australia})
ميں وی ايل ايس آئی ڈيزائن انجينيئر
(\textenglish{VLSI Design Engineer})
تھے۔ 
وہ
\textenglish{1987}
سے يونيورسٹی آف ايڈيليڈ آسٹريليا سے وابستہ ہيں جہاں وہ آجکل سکول آف اليکٹريکل اور اليکٹرانک انجينيئرنگ ميں فل پروفيسر
(\textenglish{Full Professor})
ہيں۔ ان کے نام
\textenglish{1000}
سے زيادہ تحقيقی پرچے یا اسناد حق ايجاد
(\textenglish{patents})
ہيں۔ انہيں طبيعيات کے متعدد مضامين کے ساتھ ساتھ اليکٹرانک انجينيئرنگ کی پيچيدہ نظامات
(\textenglish{complex systems})
ميں اطلاقات سے تحقیقی دلجسپی ہے ۔ وہ انسٹيٹيوٹ آف فزکس برطانيہ
(\textenglish{Institute of Physics, UK})
کے فيلو
(\textenglish{Fellow})
ہيں اور انہوں نے متعدّد ايوارڈز حاصل کئے ہيں جن ميں سال 2004 کا جنوبی آسٹريليا کا ٹال پوپی ايوارڈ
(\textenglish{South Australian Tall Poppy Award for Science})
،
سال
\textenglish{2004}
کا جنوبی آسٹريليا ميں سائينس اور ٹيکنالوجی کيلئے غير معمولی خدمات سر انجام دينے پر وزير اعلی کی طرف سے عظيم ايوارڈ
(\textenglish{Premier's SA Great Award in Science and Technology for outstanding contributions to South Australia})
،
سال
\textenglish{2012}
ميں آسٹريلوی ريسرچ کونسل کی طرف سے دی گئی فيوچر فيلوشپ
(\textenglish{Australian Research Council Future Fellowship})
،
سال
\textenglish{2015}
کا ڈيوڈ ڈيوہرسٹ ميڈل
(\textenglish{David Dewhurst Medal})
,
سال
\textenglish{2018}
کا بيری انگلس ميڈل
(\textenglish{Barry Inglis Medal})
،
اور سال
\textenglish{2019}
کا ايم اے سارجنٹ ميڈل
(\textenglish{M.~A.~Sargent Medal})
شامل ہيں۔ اس کے علاوہ وہ متعدد تحقيقی جرائد کے مدير یا ميزبان مدير ہيں اور آجکل تحقيقی جرائد
\textenglish{IEEE Access}
،
\textenglish{Frontiers in Physics}
،
\textenglish{Royal Society Open Science}
اور
\textenglish{Nature's Scientific Reports}
کے ادارياتی بورڈز کے رکن ہيں۔

\begin{thebibliography}{99}
\bibitem{Binmore}K. Binmore, Game Theory: A Very Short Introduction, Oxford University Press,~USA (2007)
\bibitem{Rasmusen}E. Rasmusen, Games \& Information: An Introduction to Game Theory, Blackwell Publishers Ltd., Oxford, 3rd Edition (2001)
\bibitem{Osborne}M. J. Osborne, An Introduction to Game Theory, Oxford University Press, USA (2003)
\bibitem{JohnNash}J. F. Nash, \textit{Non-Cooperative Games}, PhD thesis, Princeton University (1950)
\bibitem{RobertAumann}R. Aumann,~The Blackmailer Paradox: Game theory and negotiations with Arab countries (2010) . Available online at: \url{http://www.aish.com/jw/me/97755479.htm}. Accessed on 10 Dec (2018)
\bibitem{NielsenChuang}M. A. Nielsen and I. L. Chuang, \textit{Quantum Computation and Quantum Information}, Cambridge University Press (2000)
\bibitem{DavidMeyer}D. A. Meyer, Quantum strategies, Phys. Rev. Lett. 82 , 1052-1055 (1999)
\bibitem{EWL}J. Eisert, M. Wilkens and M. Lewenstein, Quantum games and quantum strategies, Phy. Rev. Lett.   83 , 3077 (1999)
\bibitem{MeyerQGQA}D. A. Meyer, Quantum games and quantum algorithms, \url{https://arxiv.org/abs/quant-ph/0004092} (2000)
\bibitem{MBroom}M. Broom, C. Cannings, and G.T. Vickers,  Choosing a nest site: Contests and catalysts, American Naturalist 147/6 , 1108-1114 (1996)
\bibitem{MaynardSmith}J. Maynard Smith, \textit{Evolution and the Theory of Games} , Cambridge University Press , ISBN 0-521-28884-3 (1982)
\bibitem{RFisher}R. A. Fisher, \textit{The Genetical Theory of Natural Selection}, Clarendon Press, Oxford (1930)
\bibitem{Weibull}J. W. Weibull, \textit{Evolutionary Game Theory}, The MIT Press, Cambridge (1995)
\bibitem{Hofbauer}J. Hofbauer and K. Sigmund, \textit{Evolutionary Games and Population Dynamics} , Cambridge University Press (1998)
\bibitem{Cressman}R. Cressman, \textit{The Stability Concept of Evolutionary Game Theory} , Springer Verlag, Berlin (1992)
\bibitem{IqbalToor2001a}A. Iqbal and A. H. Toor, Evolutionarily stable strategies in quantum games, Physics Letters A 280/5-6 , 249-256 (2001)
\bibitem{IqbalToor2001b}A. Iqbal and A. H. Toor, Entanglement and dynamic stability of Nash equilibria in a symmetric quantum game, Physics Letters A 286/4 , 245-250 (2001)
\bibitem{IqbalToor2002a}A. Iqbal and A. H. Toor, Quantum mechanics gives stability to a Nash equilibrium, Physical Review A 65 , 022306 (2002)
\bibitem{IqbalToor2002c}A. Iqbal and A. H. Toor, Darwinism in quantum systems? Physics Letters A 294/5-6 , 261-270 (2002)
\bibitem{IqbalToor2004a}Iqbal and A. H. Toor, Stability of mixed Nash equilibria in symmetric quantum games, Communications in Theoretical Physics 42/3 , 335-338 (2004)
\bibitem{AhmadNawazArxiv}Preprints of Ahmad Nawaz's publications on quantum games are available at this link:~\url{https://arxiv.org/search/quant-ph?searchtype=author&query=Nawaz%2C+A}
\bibitem{MRamzanArxiv}Preprints of Mohammad Ramzan's publications on quantum games are available at this link: \url{https://arxiv.org/search/quant-ph?searchtype=author&query=Ramzan%2C+M}
\bibitem{SalmanKhanArxiv}Preprints of Salman Khan's publications on quantum games are available at this link: \url{https://arxiv.org/search/quant-ph?searchtype=author&query=Khan%2C+S}
\bibitem{FaisalShahKhan}Preprints of Faisal Shah Khan's publications on quantum games are available at this link:
\url{https://arxiv.org/search/quant-ph?searchtype=author&query=Khan%2C+F+S}
\bibitem{AzharArxiv}Preprints of Azhar Iqbal's publications on quantum games are available at this link: \url{https://arxiv.org/search/quant-ph?searchtype=author&query=Iqbal%2C+A}
\bibitem{BenjaminHaydenComment} S. C. Benjamin and P. M. Hayden, Comment on ``Quantum Games and Quantum Strategies'', Physical Review Letters \textbf{87}, 069801 (2001). DOI: 10.1103/PhysRevLett.87.069801.
\bibitem{IqbalAbbott2026} A. Iqbal and D. Abbott, ``Evolutionary stability and payoff equivalence in an Eisert--Wilkens--Lewenstein quantum game'', Quantum Information Processing \textbf{25}, 348 (2026). DOI: 10.1007/s11128-026-05373-0.
\bibitem{EisertReply} J. Eisert, M. Wilkens and M. Lewenstein, ``Eisert, Wilkens, and Lewenstein Reply,'' Physical Review Letters \textbf{87}, 069802 (2001). DOI: 10.1103/PhysRevLett.87.069802.
\bibitem{BenjaminHaydenMultiplayer} S. C. Benjamin and P. M. Hayden, ``Multiplayer quantum games,'' Physical Review A \textbf{64}, 030301(R) (2001). DOI: 10.1103/PhysRevA.64.030301.
\bibitem{KhanReview2018} F. S. Khan, N. Solmeyer, R. Balu and T. S. Humble, ``Quantum games: a review of the history, current state, and interpretation,'' Quantum Information Processing \textbf{17}, 309 (2018). DOI: 10.1007/s11128-018-2082-8.
\bibitem{AzharChapter} A. Iqbal and T. Cheon, Evolutionary stability in quantum games, Chapter 13 in Quantum Aspects of Life, edited by D. Abbott, P.C.W. Davies and A. K. Pati, foreword by Sir Roger Penrose, ISBN 978-1-84816-267-9 , Imperial College Press, preprint: \url{https://arxiv.org/abs/0706.1413} (2008)
\end{thebibliography}
\end{document}